\documentclass[aps,prd,preprint,12pt,letterpaper]{revtex4-2}
\usepackage{cancel}
\usepackage{bm}
\usepackage{tabularx}
\usepackage{amsmath, amsfonts, amssymb}
\usepackage{braket}
\usepackage{graphicx}
\usepackage{enumitem}
\usepackage[normalem]{ulem}
\usepackage{subcaption}
\usepackage{xcolor}
\usepackage[final]{hyperref}
\hypersetup{
    colorlinks=true,
    linkcolor=blue,
    citecolor=blue,
    filecolor=magenta,
    urlcolor=blue
}

\newcommand{\aabs}{\alpha_{\rm abs}}
   
\begin{document}
\raggedbottom
\title{Combining Pumping and Absorption Effects in Neutrino Oscillation
Probabilities: A Non-Hermitian Framework}

\author{Gabriela Barenboim}
\email{gabriela.barenboim@uv.es}
\affiliation{Departament de Física Teòrica, Universitat de València, C/ Dr. Moliner 50, Burjassot 46100, Spain}
\affiliation{Instituto de Física Corpuscular, CSIC-Universitat de València, C/Catedrático José Beltrán 2, Paterna 46980, Spain}

\author{Alberto M. Gago}
\email{agago@pucp.edu.pe}
\author{Eduardo Massoni} \email{emassoni@pucp.edu.pe}
\affiliation{Sección Física, Departamento de Ciencias, Pontificia Universidad Cat\'olica del Per\'u, Apartado 1761, Lima, Per\'u}

\date{\today}
\begin{abstract}
We study two-flavour neutrino oscillations in which a coherent
off-diagonal interaction from a complex background field coexists with an
absorptive channel acting on a single mass eigenstate. Since the two do
not commute, the effective Hamiltonian is non-normal and oscillations and
damping no longer factorise. We formulate the dynamics both as the no-jump sector 
of a Lindblad
evolution and through biorthogonal diagonalisation, and obtain the
evolution operator of our model in closed form. The model interpolates
between quasi-global damping, standard invisible decay, and a critical
point with polynomial-times-exponential evolution; for initially pure
states it describes norm loss rather than decoherence. Comparing
neutrinos and antineutrinos in the same background, with the mixing
phase $\phi$ reversed and the background phase $\delta$ fixed, we derive
closed-form $CP$, $CPT$ and $T$ asymmetries. The $CPT$ asymmetries arise
only from this phase assignment. In contrast, the $T$ asymmetry, which is
forbidden in unitary two-flavour oscillations, is independent of the
prescription: it arises only through the non-normal structure, is
proportional to the absorptive coupling, and vanishes only when
$\delta+\phi=n\pi$. For
equal initial populations the net neutrino--antineutrino asymmetry
vanishes identically, while a non-zero flavour asymmetry survives and
requires a non-normal Hamiltonian.

\end{abstract}
\maketitle


\section{Introduction}
\label{sec:intro}
Current neutrino oscillation data are very well described by the misalignment between neutrino flavour eigenstates and mass eigenstates, together with the quantum interference generated by nonzero neutrino mass-squared differences~\cite{DayaBay:2022orm,T2K:2025wet}. Although mass-induced oscillations provide the established leading framework, subleading effects associated with physics beyond standard oscillations, such as non-standard neutrino interactions~\cite{Valle:1987gv,Guzzo:1991hi}, Lorentz-invariance violation~\cite{Kostelecky:2003cr,Kostelecky:2003xn}, neutrino decay~\cite{Gelmini:1983ea,Schechter:1981cv}, or quantum decoherence~\cite{Benatti:2000ph,Benatti:2001fa}, among others~\cite{Gasperini:1988zf,Barbieri:2000mg}, may also influence neutrino propagation. These possibilities have been extensively studied, mostly by adding one new-physics ingredient at a time to the standard oscillation Hamiltonian. 
A theoretically interesting and comparatively less explored situation arises when two distinct beyond-standard mechanisms coexist and compete with the standard oscillation dynamics. In particular, the simultaneous presence, in the neutrino mass basis, of a coherent off-diagonal interaction and a diagonal selective absorptive or decay channel can lead to a non-Hermitian and non-normal effective Hamiltonian, opening the possibility of qualitatively new phenomenology.

Non-Hermitian Hamiltonians naturally appear in neutrino oscillation physics when invisible decay or absorptive effects are included~\cite{Berryman:2014yoa,Lindner:2001fx}. If the Hermitian and anti-Hermitian components are simultaneously diagonalizable, the Hamiltonian is normal and the evolution reduces to the familiar picture of oscillations modulated by exponential damping. However, when these two components do not commute, the Hamiltonian becomes non-normal, and the oscillatory and absorptive parts of the evolution cannot be factorized in a trivial way. A detailed analytical treatment of this situation in the context of neutrino decay was developed in~\cite{Chattopadhyay:2021eba}, where the decay matrix acquires off-diagonal components in the neutrino mass basis. 

In the present work, we explore a complementary realization. Unlike the decay-mismatch setup mentioned above, where non-normality is associated with an off-diagonal anti-Hermitian component in the mass basis, here the off-diagonal structure is complex, coherent, and Hermitian, while the absorptive channel is diagonal and selective. The coherent off-diagonal term is not tied to a unique microscopic origin. As discussed in~\cite{barenboim3},  such a structure may arise from violation of the equivalence principle~\cite{Diaz:2020aax} or from a CPT-odd interaction sector~\cite{Barger:2000iv}, whenever the corresponding interaction eigenbasis is misaligned with the neutrino mass basis. Analogous complex coherent contributions may also be induced by off-diagonal non-standard interactions in matter~\cite{Farzan:2017xzy} or by a classical axion-like background field~\cite{Huang:2018cwo}. 

We therefore adopt a phenomenological approach and describe this contribution as a complex coherent pumping interaction coexisting with a selective absorptive/decay channel. In order to isolate the essential physical effects, we restrict our study to a two-flavour framework. This simplification is not merely pedagogical: even in the two-flavour case, the interplay between standard oscillations, coherent pumping, and selective loss gives rise to a non-Hermitian and non-normal Hamiltonian whose phenomenology is already rich. In particular, we focus on the $CP$ and $CPT$ asymmetric effects that arise when the intrinsic complex phase of the pumping interaction is assumed to remain invariant under the corresponding $CP/CPT$ transformations. Using the analytical expressions derived in this framework, we perform an exploratory analysis of whether the resulting active-neutrino lepton asymmetry could provide a viable source for the observed baryon asymmetry after electroweak sphaleron conversion~\cite{fukugita_yanagida}. 

As emphasized above, extracting phenomenological consequences from non-normal systems is a nontrivial task, since the coherent and absorptive parts of the evolution cannot, in general, be factorized.A further methodological contribution of this work is to 
show how this
class of non-Hermitian systems can be analysed both within the
density-matrix formalism and through biorthogonal diagonalisation of the
Hamiltonian. The explicit solution presented here is obtained in closed
form from the Cayley--Hamilton theorem, while the density-matrix
formulation provides a natural bridge toward a more general open-system
description.

The rest of this paper is organised as follows.
Section~\ref{sec:formalism} establishes the mathematical framework,
covering both the density-matrix and Hamiltonian formulations and
introducing the biorthogonal basis needed for the analytic solution.
  Section~\ref{sec:model} introduces the specific non-Hermitian,
  non-normal model, derives the evolution operator in closed form,
  discusses possible microscopic interpretations in terms of coherent
  background fields and absorptive/decay effects, and maps out the
  corresponding phenomenological regimes. The phase prescription
  relating neutrinos and antineutrinos and the resulting $CP$, $T$ and
  $CPT$ asymmetries are presented in Section~\ref{sec:asymmetries}.
  Section~\ref{sec:application} discusses the purity structure of the
  evolution and the implications for neutrino flavour asymmetries and
  leptogenesis.
We conclude in Section~\ref{sec:conclusions}.

\section{Non-Hermitian Neutrino Dynamics: Density-Matrix
and Hamiltonian Approaches}
\label{sec:formalism}

In this section, we establish the theoretical framework for non-Hermitian neutrino dynamics, introducing both the density matrix and Hamiltonian formulations and identifying the mathematical conditions that distinguish normal from non-normal regimes.
\subsection{Non-Hermitian Hamiltonians and operator normality}

A non-Hermitian Hamiltonian can always be decomposed as
\begin{equation}
H_{\mathrm{NH}} = H_0 - \frac{i}{2}\Gamma,
\label{nonHerH}
\end{equation}
with $H_0$ and $\Gamma$ Hermitian. The mathematical properties of
$H_{\mathrm{NH}}$ determine which computational approach is valid~\cite{HornJohnson2013}:

\begin{enumerate}[label=(\alph*)]
\item \textbf{Normal} ($[H_0,\Gamma]=0$): $H_{\mathrm{NH}}$ commutes
with its adjoint, $[H_{\mathrm{NH}},H_{\mathrm{NH}}^\dagger]=0$.
Its eigenvectors form an orthonormal basis and the diagonalisation
satisfies $U^{-1}=U^\dagger$.

\item \textbf{Non-normal} ($[H_0,\Gamma]\neq 0$): $H_{\mathrm{NH}}$
cannot be diagonalised by a unitary transformation, and its
eigenvectors do not, in general, form an orthonormal basis. The
biorthogonal framework developed below is required.
\end{enumerate}

The normality condition, $[H_0,\Gamma]=0$, is trivially satisfied when both $H_0$ and $\Gamma$ are diagonal in the same basis. When $\Gamma$ is diagonal and $H_0$ contains off-diagonal elements, normality is preserved only if the Hermitian dynamics generated by $H_0$ acts within subspaces where the absorptive coefficients (i.e., the eigenvalues of $\Gamma$) are equal. In other words, $H_0$ must not couple states with different absorptive strengths. Otherwise, the full Hamiltonian is non-normal.

In the more general case where both $H_0$ and $\Gamma$ are non-diagonal, normality is achieved if and only if they commute, which is equivalent to requiring that they can be simultaneously diagonalized.

Standard treatments in the literature, such as neutrino decay and global (overall) damping, typically correspond to normal Hamiltonians, where absorption either acts uniformly or remains aligned with the mixing structure. In contrast, our framework focuses on the non-normal case, which gives rise to qualitatively new phenomenology.

\subsection{Density-matrix formulation of non-Hermitian evolution}

\subsubsection{The non-Hermitian Liouville equation}

The non-Hermitian Liouville evolution equation of the density matrix is obtained from the
Schr\"odinger equation
\begin{equation}
i\,\frac{d}{dt}\ket{\psi} = H_{\mathrm{NH}}\ket{\psi}
\label{Schrod}
\end{equation}
together with $\rho=\ket{\psi}\bra{\psi}$. This yields the
non-trace-preserving equation
\begin{equation}
{\mathcal{L}_{\rm NH}}[\rho(t)]
= -i\!\left(H_{\mathrm{NH}}\rho(t) - \rho(t)H_{\mathrm{NH}}^\dagger\right).
\label{eq:LiouvilleNH}
\end{equation}
where ${\mathcal{L}_{\rm NH}}[\rho(t)]=\dot\rho(t)$. Now,
inserting the decomposition \eqref{nonHerH} this becomes
\begin{equation}
{\mathcal{L}_{\rm NH}}[\rho(t)]
= -i[H_0,\rho(t)] - \tfrac{1}{2}\{\Gamma,\rho(t)\},
\label{genLind2}
\end{equation}
which is completely general and does not require $H_{\mathrm{NH}}$
to be Hermitian, normal, or diagonalisable. 
The anti-commutator term drives trace, or norm, loss. Therefore,
Eq.~(\ref{genLind2}) should be understood as the effective non-Hermitian,
or no-jump, part of a Lindblad evolution. Comparing Eq.~(\ref{genLind2})
with the trace-preserving Lindblad equation, our loss term can be identified
with the Lindblad anti-commutator term
\(-\frac{1}{2}\{L_k^\dagger L_k,\rho\}\), or more generally with
\(-\frac{1}{2}\sum_k\{L_k^\dagger L_k,\rho\}\)~\cite{BreuerPetruccione2002}. Equivalently, one may identify
\(\Gamma\) with \(\sum_k L_k^\dagger L_k\). Here \(L_k\) denote the Lindblad operators. The corresponding jump terms,
\(L_k\rho L_k^\dagger\), which would restore the trace, are absent in
Eq.~(\ref{genLind2}). For comparison, a recent open-quantum-system
treatment of visible neutrino decay, including regeneration terms, has
been formulated in terms of the Lindblad equation~\cite{Kopp:2026tnx}.

\subsubsection{Biorthogonal basis and Liouvillian eigenoperators}
\label{sec:biorthogonal}

When $H_{\mathrm{NH}}$ is diagonalisable, Eq.~\eqref{eq:LiouvilleNH}
admits a closed analytic solution. We assume this henceforth; it is
not required for the general validity of Eq.~\eqref{genLind2}.

The right and left eigenvectors are defined by~\cite{Brody2014}
\begin{equation}
H_{\mathrm{NH}}\,|\phi_n^R\rangle = \lambda_n|\phi_n^R\rangle,
\qquad
\langle\phi_n^L|\,H_{\mathrm{NH}} = \lambda_n\langle\phi_n^L|,
\quad \lambda_n\in\mathbb{C}.
\end{equation}
They satisfy the biorthogonality condition
\begin{equation}
\langle\phi_m^L|\phi_n^R\rangle=\delta_{mn}, 
\label{biortho}
\end{equation}
which guarantees the resolution of the identity
$\sum_{n}|\phi_n^R\rangle\langle\phi_n^L|=\mathbb{I}$,
so that any state vector expands as
\(\ket{\psi(t)}=\sum_m c_m(t)\ket{\phi_m^R}\).

Therefore, a pure-state density matrix, \(\rho =\ket{\psi} \bra{\psi}   \), can be written~\cite{Niu2023}
\begin{equation}
\rho(t)
=
\sum_{m,n}
\rho_{mn}(t)
|\phi_m^R\rangle\langle\phi_n^R|,
\label{rhoexpand}
\end{equation}
where $\rho_{mn} =c_m c^*_n$. Then, replacing Eq.(\ref{rhoexpand})
into the Liouville
equation Eq.(\ref{eq:LiouvilleNH}):
\begin{equation}
\mathcal{L}_{\rm NH}[\rho(t)]
=\sum_{m,n}\dot\rho_{mn}(t)|\phi_m^R\rangle\langle\phi_n^R|
=-i \sum_{m,n}(\lambda_m-\lambda_n^*)\rho_{mn}\,|\phi_m^R\rangle\langle\phi_n^R|,
\label{eq:liouvillian_eigenoperators}
\end{equation}
Consequently, the above
equation can be decoupled into independent equations for each component:
\begin{equation}
\dot\rho_{mn}(t)
= -i(\lambda_m-\lambda_n^*)\,\rho_{mn}(t),
\label{eq:rho_mn_evolution}
\end{equation}
each solved by a simple exponential, giving a complete analytic
solution.

\subsubsection{Vectorized representation in the eigenoperator basis}

We restrict ourselves to the two-flavour case, which captures the essential
features of the construction. The extension to three flavours is
straightforward, provided that \(H_{\rm NH}\) remains diagonalisable.

In the two-flavour case, we vectorize the density matrix by arranging
the coefficients \(\rho_{mn}\) in the eigenoperator basis into a column
vector~\cite{Gilchrist2009}. Explicitly, we define
\[
\boldsymbol{\varrho}_{\rm D}
=
\left(
\rho_{11},
\rho_{12},
\rho_{21},
\rho_{22}
\right)^T .
\]
Thus, \(\boldsymbol{\varrho}_{\rm D}\) denotes the vectorized form of
\(\rho\) in the eigenoperator basis. Equivalently, its components may be
written as \(\varrho^r_{\rm D}=\rho_{mn}\), where the single index
\(r=0,1,2,3\) labels the ordered pairs
\((m,n)=(1,1),(1,2),(2,1),(2,2)\), respectively. With this convention,
the Liouville equation becomes
\begin{equation}
\dot{\boldsymbol{\varrho}}_{\rm D}
=
\mathbb L_{\rm D}
\boldsymbol{\varrho}_{\rm D},
\end{equation}
where
\[
\mathbb L_{\rm D}
=
-i\,{\rm diag}
\left(
\lambda_1-\lambda_1^*,
\lambda_1-\lambda_2^*,
\lambda_2-\lambda_1^*,
\lambda_2-\lambda_2^*
\right).
\]
The solution is therefore
\begin{equation}
\boldsymbol{\varrho}_{\rm D}(t)
=
e^{\mathbb L_{\rm D}t}
\boldsymbol{\varrho}_{\rm D}(0).
\end{equation}

\subsubsection{Pauli-basis representation}

We now introduce a second operator basis. The same density matrix \(\rho\)
can be expanded in the Pauli operator basis as~\cite{BreuerPetruccione2002}
\begin{equation}
\rho
=
\frac12
\sum_{w=0}^{3}
\varrho^w_{\rm P}\sigma_w,
\qquad
\varrho^w_{\rm P}
=
{\rm Tr}(\sigma_w\rho),
\end{equation}
with \(\sigma_0=\mathbb I\), and
\[
\boldsymbol{\varrho}_{\rm P}
=
(\varrho^0_{\rm P},\varrho^1_{\rm P},
 \varrho^2_{\rm P},\varrho^3_{\rm P})^T .
\]
The vectors \(\boldsymbol{\varrho}_{\rm D}\) and
\(\boldsymbol{\varrho}_{\rm P}\) are two different vector representations
of the same density matrix: the former is associated with the Liouvillian
eigenoperator basis, whereas the latter is associated with the Pauli basis~\cite{Gilchrist2009}.
The transformation between them is given by
\begin{equation}
\varrho^w_{\rm P}
=
\sum_{m,n} S_{w,(mn)}\rho_{mn}
=
\sum_{r} S_{wr}\varrho^r_{\rm D},
\label{strans}
\end{equation}
where
\[
S_{w,(mn)}
=
{\rm Tr}\!\left(
\sigma_w|\phi_m^R\rangle\langle\phi_n^R|
\right),
\]
and \(S_{wr}\equiv S_{w,(mn)}\) under the ordering
\(r=0,1,2,3 \leftrightarrow (11),(12),(21),(22)\). In compact form,
Eq.~\eqref{strans} reads
\begin{equation}
\boldsymbol{\varrho}_{\rm P}
=
S\boldsymbol{\varrho}_{\rm D},
\end{equation}
where \(S\) is a \(4\times4\) matrix in the two-flavour framework. 
The evolution in the Pauli basis is therefore
\begin{equation}
\dot{\boldsymbol{\varrho}}_{\rm P}
=
\mathbb L_{\rm P}\boldsymbol{\varrho}_{\rm P},
\qquad
\mathbb L_{\rm P}
=
S\mathbb L_{\rm D}S^{-1}.
\label{liouPauli}
\end{equation}
Equivalently,
\begin{equation}
\boldsymbol{\varrho}_{\rm P}(t)
=
M(t)\boldsymbol{\varrho}_{\rm P}(0),
\qquad
M(t)
=
e^{\mathbb L_{\rm P}t}
=
S e^{\mathbb L_{\rm D}t}S^{-1}.
\end{equation}
It is useful to make explicit how the Pauli-basis generator
\(\mathbb L_{\rm P}\) is obtained from the non-Hermitian Liouvillian
written in Eq.~\eqref{genLind2}. In components, the first relation in
Eq.~\eqref{liouPauli} reads
\begin{equation}
\dot{\varrho}_{\rm P}^{\mu}
=
\sum_{\nu=0}^{3}
(\mathbb L_{\rm P})_{\mu\nu}\,
\varrho_{\rm P}^{\nu},
\label{LPcomponents}
\end{equation}
where the matrix elements are given by
\begin{equation}
(\mathbb L_{\rm P})_{\mu\nu}
=
\frac{1}{2}
{\rm Tr}\!\left[
\sigma_\mu\,
\mathcal L_{\rm NH}[\sigma_\nu]
\right].
\label{LPmatrixelements}
\end{equation}
Thus, \(\mathbb L_{\rm P}\) is the matrix representation of the
non-Hermitian Liouvillian \(\mathcal L_{\rm NH}\) in the Pauli operator
basis. The derivation of Eq.~\eqref{LPmatrixelements} is given in
Appendix~\ref{app:PauliLiouvillian}.

\subsubsection{Effective flavour-transition probability}

The flavour transition probability is defined as
\begin{equation}
P_{\nu_\alpha\to\nu_\beta}(t)
= \mathrm{Tr}\!\left(\rho_{\nu_\beta}(0)\,\rho_{\nu_\alpha}(t)\right)
= \tfrac{1}{2}\sum_{w,\eta=0}^3
{\varrho}^w_{{\nu_\beta}\rm P}
    \,M_{w\eta}(t)\,
  {\varrho}^\eta_{{\nu_\alpha}\rm P}(0).
\label{probrho}
\end{equation}
It is worth noting that in the non-Hermitian case probability is not
conserved, and therefore the indices in Eq.~\eqref{probrho} run from
zero. This is in contrast to the Lindblad case, where the evolution
is trace-preserving and the $w=0$ component is trivially constant.

\subsection{Hamiltonian formulation}

While the density-matrix formulation provides a basis-independent
description, the Schr\"odinger formulation allows a direct and
operational computation of transition amplitudes once the biorthogonal
diagonalisation is performed.

Assuming that \(H_{\mathrm{NH}}\) is diagonalisable in the neutrino mass
basis, we write
\[
H_{\mathrm{NH}}=P\,D\,P^{-1},
\qquad
D=\mathrm{diag}(\lambda_1,\lambda_2,\ldots),
\]
where the columns of \(P\) are the right eigenvectors of
\(H_{\mathrm{NH}}\), while the rows of \(P^{-1}\) are the corresponding
left eigenvectors, normalized according to the biorthogonality condition (see Eq.~(\ref{biortho})).

The matrices \(P\) and \(P^{-1}\) satisfy \(P^{-1}P=\mathbb I\), but in
general \(P^{-1}\neq P^\dagger\). Therefore,
\[
e^{-iH_{\mathrm{NH}}t}
=
P\,e^{-iDt}\,P^{-1}.
\]
A related application of such a non-unitary diagonalisation appears in
the treatment of neutrino decay in matter, where the effective
Hamiltonian has complex eigenvalues and is diagonalised by a
non-unitary mixing matrix~\cite{Ascencio-Sosa:2018lbk}. Then, introducing the effective non-unitary mixing matrices
$\widetilde{U}\equiv UP$ and $\widetilde{U}^{-1}\equiv P^{-1}U^\dagger$, where $U$ is the neutrino mixing matrix, 
the flavour transition amplitude takes the compact form
\begin{equation}
A_{\alpha\to\beta}(t)
= \bigl[U\,P\,e^{-iDt}\,P^{-1}U^\dagger\bigr]_{\beta\alpha}
= \sum_{i,j,n}
    U_{\beta j}\,P_{jn}\,e^{-i\lambda_n t}\,(P^{-1})_{ni}\,U_{\alpha i}^*
= \sum_n
    \widetilde{U}_{\beta n}\,(\widetilde{U}^{-1})_{n\alpha}\,
    e^{-i\lambda_n t}.
\label{eq:Amp_compact}
\end{equation}

\subsubsection{Non-unitary two-flavour oscillation probability}

The corresponding transition probability is
$P_{\nu_\alpha\to\nu_\beta}(t)=|A_{\alpha\to\beta}(t)|^2$.
Defining $\lambda_{12}\equiv\lambda_1-\lambda_2$ and
$A\equiv\widetilde{U}_{\beta1}\widetilde{U}_{\beta2}^*
(\widetilde{U}^{-1})_{1\alpha}(\widetilde{U}^{-1})_{2\alpha}^*$,
the two-flavour probability reads
\begin{align}
P_{\nu_\alpha\to\nu_\beta}(L)
&= \left|(\widetilde{U}^{-1})_{1\alpha}\right|^2
   \left|\widetilde{U}_{\beta1}\right|^2 e^{2\,\mathrm{Im}(\lambda_1)L}
+  \left|(\widetilde{U}^{-1})_{2\alpha}\right|^2
   \left|\widetilde{U}_{\beta2}\right|^2 e^{2\,\mathrm{Im}(\lambda_2)L}
\nonumber\\
&\quad
+ 2\,e^{\mathrm{Im}(\lambda_1+\lambda_2)L}
  \Big[\mathrm{Re}(A)\cos\!\bigl(\mathrm{Re}(\lambda_{12})L\bigr)
  +\mathrm{Im}(A)\sin\!\bigl(\mathrm{Re}(\lambda_{12})L\bigr)\Big].
\label{eq:prob_twoflavor}
\end{align}
This formula reveals the key features of our framework:
\begin{itemize}
\item \textbf{Exponential damping:} The factors $e^{2\mathrm{Im}(\lambda_i)L}$
    suppress the individual mass-eigenstate contributions.
\item \textbf{Modified oscillations:} The oscillatory terms have
    frequencies determined by $\mathrm{Re}(\lambda_{12})$.
\item \textbf{Interference:} The cross-term combines both effects,
    with its amplitude also exponentially damped.
\end{itemize}
When $H_{\rm NH}$ is Hermitian this reduces to the standard unitary result.
\section{The Model: Non-Hermitian and Non-Normal Hamiltonian Structure}
\label{sec:model}

\subsection{The Hamiltonian}

We consider a two-flavour neutrino system described by the effective
non-normal Hamiltonian
\begin{equation}
H_{\mathrm{eff}}=
\begin{pmatrix}
\Delta & \lambda e^{i\delta}\\
\lambda e^{-i\delta} & -\Delta-i\aabs
\end{pmatrix},
\label{HeffGRAL}
\end{equation}
where:
\begin{itemize}
 \item $\Delta=(m_1^2-m_2^2)/(4E_\nu)$ is the standard oscillation
    parameter; since the absorptive term acts on the heavier state
    $\nu_2$, $\Delta<0$,
\item $\lambda$ represents the pumping strength of the off-diagonal
    coupling,
\item $\delta$ is the complex phase of the off-diagonal coherent background field with magnitude $\lambda$,
\item $\aabs$ parametrises the absorption/decay rate.
\end{itemize}
The Hamiltonian decomposes as $H_{\mathrm{eff}}=H_{\mathrm{herm}}-i\Gamma/2$
with
\begin{equation}
H_{\mathrm{herm}} =
\begin{pmatrix}\Delta & \lambda e^{i\delta}\\
\lambda e^{-i\delta} & -\Delta\end{pmatrix},
\qquad
\Gamma =
\begin{pmatrix}0 & 0\\ 0 & 2\aabs\end{pmatrix}.
\end{equation}
\textbf{Crucial feature:} $[H_{\mathrm{herm}},\Gamma]\neq 0$.
This non-commuting structure distinguishes our framework from standard
approaches and is responsible for the rich phenomenology we explore.

\subsubsection{Possible microscopic illustrations of coherent background fields and absorptive effects}

The effective Hamiltonian \eqref{HeffGRAL} provides a compact
phenomenological parametrisation of scenarios in which a coherent
Hermitian contribution drives flavour mixing through the off-diagonal
coupling $\lambda e^{\pm i\delta}$, while an absorptive component
induces loss into unobserved channels through the anti-Hermitian term
$-i\aabs$. In this work we remain agnostic about the microscopic
origin of these terms. Nevertheless, several possible realizations can
motivate this structure. The Hermitian sector may arise, for instance,
from beyond-standard coherent physics, such as non-standard neutrino
interactions in matter, violations of the equivalence principle,
$CPT$-odd interaction sectors, or coherent classical background fields.
The absorptive sector, on the other hand, may represent invisible
decay, absorption, or effective loss into degrees of freedom that are
not observed in the active-neutrino sector.

At the two-state effective level, and after an appropriate choice of
basis, a broad class of microscopic mechanisms leading to coherent
mixing together with trace loss can be parametrized by the structure of
Eq.~\eqref{HeffGRAL}.

\textbf{Axion-like background field:}
A coherent axion-like background field provides one possible
illustration~\cite{Huang:2018cwo}. If the field can be treated as a
classical external background, its effect on neutrino propagation may be
described in analogy with a generalized matter potential. In the
neutrino mass basis, such a contribution can be effectively represented
by off-diagonal terms of the form \(\lambda e^{\pm i\delta}\) in the
Hamiltonian. In this interpretation, the phase \(\delta\) encodes the
intrinsic phase of the coherent background interaction, while
\(\lambda\) measures its effective strength.

The anti-Hermitian term $-i\aabs$, instead, parametrizes a selective
absorptive or decay process leading to flux loss into unobserved
channels. As shown below, depending on the relative size of
$\lambda$, $\Delta$, and $\aabs$, the system can approach regimes
that are phenomenologically close to invisible neutrino decay or to
neutrino loss.

\subsection{Solving the system}
\label{solving}

A useful decomposition is
\begin{equation}
H_{\mathrm{eff}} = -i\frac{\aabs}{2}\,\mathbb{I}
+ \underbrace{
\begin{pmatrix}
\Delta+i\frac{\aabs}{2} & \lambda e^{i\delta}\\
\lambda e^{-i\delta} & -\Delta-i\frac{\aabs}{2}
\end{pmatrix}}_{\mathcal{M}},
\end{equation}
with eigenvalues
\begin{equation}
E_\pm = -\frac{i\aabs}{2}
\pm\sqrt{\left(\Delta^2+\lambda^2-\frac{\aabs^2}{4}\right)
    +i\,\aabs\Delta}\,.
\label{Enereigenvalues}
\end{equation}
Since $\mathcal{M}$ is traceless, $\mathcal{M}^2=\Lambda^2\mathbb{I}$
(Cayley--Hamilton), and the evolution operator takes the closed form 
\begin{equation}
\mathcal{U}(t) = e^{-\frac{\aabs}{2}t}
\!\left(\cos(\Lambda t) - i\sin(\Lambda t)\frac{\mathcal{M}}{\Lambda}\right),
\label{evolOperator}
\end{equation}
where 
\begin{equation}
\Lambda \equiv
\sqrt{\left(\Delta^2+\lambda^2-\frac{\aabs^2}{4}\right)
    +i\,\aabs\Delta}
\;=\; w+ik,
\label{eq:Lambda}
\end{equation}
with $w=\mathrm{Re}[\Lambda]$ and $k=\mathrm{Im}[\Lambda]$. 

The elements of the evolution operator matrix $\mathcal{U}(t)$ are given by: 
\begin{align}
u_{11} &= e^{-\frac{\aabs}{2}t}\!\left(
    \cos(\Lambda t)
    - i\frac{\Delta+i\frac{\aabs}{2}}{\Lambda}\sin(\Lambda t)\right),
\qquad
u_{22} = e^{-\frac{\aabs}{2}t}\!\left(
    \cos(\Lambda t)
    + i\frac{\Delta+i\frac{\aabs}{2}}{\Lambda}\sin(\Lambda t)\right),
  \label{u11u22}  \\
u_{12} &= -i\,e^{-\frac{\aabs}{2}t}
    \frac{\lambda e^{i\delta}}{\Lambda}\sin(\Lambda t),\qquad
u_{21} = -i\,e^{-\frac{\aabs}{2}t}
    \frac{\lambda e^{-i\delta}}{\Lambda}\sin(\Lambda t).
    \label{u12}
\end{align}

\subsubsection{Neutrino Oscillation Probability Structure}

For calculating the neutrino oscillation probability, we need to define the amplitude matrix $A$:
\begin{equation}
A = \begin{pmatrix}
A_{ee} & A_{e\mu}\\
A_{\mu e} & A_{\mu\mu}
\end{pmatrix}=R(\theta,\phi)\,\mathcal{U}(t)\,R^\dagger(\theta,\phi),
\qquad
R(\theta,\phi) =
\begin{pmatrix}
\cos\theta & \sin\theta\,e^{-i\phi}\\
-\sin\theta\,e^{i\phi} & \cos\theta
\end{pmatrix}.
\end{equation}
Each oscillation probability has the general Fourier structure
\begin{equation}
P_{\alpha\beta}\equiv P_{\nu_\alpha\to\nu_\beta}=|A_{\beta\alpha}|^2= A_0 + A_c\cos\phi + A_s\sin\phi
+ A_{2c}\cos 2\phi + A_{2s}\sin 2\phi.
\end{equation}
Next we present these coefficients for the survival and transition probability.
\begin{itemize}
    \item {\bf Survival probability coefficients for $P_{ee}$ and $P_{\mu\mu}$}

From the amplitude $A_{ee}$
    \begin{equation}
A_{ee} = u_{11}\cos^2\theta + u_{22}\sin^2\theta
+ \sin\theta\cos\theta\,(u_{12}e^{i\phi}+u_{21}e^{-i\phi}),
\label{Aee}
\end{equation}
we obtain the following coefficients:
\begin{align}
A^{ee}_0 &= |u_{11}|^2\cos^4\theta + |u_{22}|^2\sin^4\theta
    + \sin^2\theta\cos^2\theta\,(|u_{12}|^2+|u_{21}|^2)
    + 2\sin^2\theta\cos^2\theta\,\mathrm{Re}(u_{11}u_{22}^*),\\
A^{ee}_c &= 2\sin\theta\cos\theta\!\left(
    \cos^2\theta\,\mathrm{Re}(u_{11}u_{12}^*+u_{11}u_{21}^*)
    +\sin^2\theta\,\mathrm{Re}(u_{22}u_{12}^*+u_{22}u_{21}^*)\right),\\
A^{ee}_s &= -2\sin\theta\cos\theta\!\left(
    \cos^2\theta\,\mathrm{Im}(u_{11}u_{21}^*-u_{11}u_{12}^*)
    +\sin^2\theta\,\mathrm{Im}(u_{22}u_{21}^*-u_{22}u_{12}^*)\right),\\
A^{ee}_{2c} &= 2\sin^2\theta\cos^2\theta\,\mathrm{Re}(u_{12}u_{21}^*),\quad
A^{ee}_{2s} = -2\sin^2\theta\cos^2\theta\,\mathrm{Im}(u_{12}u_{21}^*).
\end{align}
On the other hand, amplitude $A_{\mu\mu}$ is given by
\begin{equation}
A_{\mu\mu} = u_{22}\cos^2\theta + u_{11}\sin^2\theta
- \sin\theta\cos\theta\,(u_{12}e^{i\phi}+u_{21}e^{-i\phi}).
\label{Auu}
\end{equation}
Thus, from the comparison between Eq.(\ref{Aee}) and Eq.(\ref{Auu}), it is straightforward to obtain the coefficients for $P_{\mu\mu}$ by applying the transformations $u_{11}\leftrightarrow u_{22}$ and $\theta\to -\theta$.

\item{\bf Transition probability coefficients for $P_{e \mu}$ and $P_{\mu e}$}

The amplitude $A_{e \mu}$ is defined by: 
\begin{equation}
A_{e\mu} = u_{12}\cos^2\theta
- e^{-i\phi}\sin\theta\cos\theta\,b
- e^{-2i\phi}u_{21}\sin^2\theta.
\label{Aemu}
\end{equation}
The corresponding coefficients of $P_{\mu e}=|A_{e\mu}|^2$ are:
\begin{align}
A_0 &= |u_{12}|^2\cos^4\theta + |u_{21}|^2\sin^4\theta
    + |b|^2\sin^2\theta\cos^2\theta,\\
A_c &= 2\sin\theta\cos\theta\!\left(
    \sin^2\theta\,\mathrm{Re}(u_{21}b^*)
    -\cos^2\theta\,\mathrm{Re}(u_{12}b^*)\right),\\
A_s &= 2\sin\theta\cos\theta\!\left(
    \sin^2\theta\,\mathrm{Im}(u_{21}b^*)
     +\cos^2\theta\,\mathrm{Im}(u_{12}b^*)\right),\\     
A_{2c} &= -2\sin^2\theta\cos^2\theta\,\mathrm{Re}(u_{12}^*u_{21}),\quad
A_{2s} = - 2\sin^2\theta\cos^2\theta\,\mathrm{Im}(u_{12}^*u_{21}), 
\end{align}
where $b\equiv u_{11}-u_{22}$. Meanwhile, the 
amplitude $A_{\mu e}$ is:
\begin{equation}
A_{\mu e} = u_{21}\cos^2\theta
- e^{i\phi}\sin\theta\cos\theta\,b
- e^{2i\phi}u_{12}\sin^2\theta.
\label{Amue}
\end{equation}
As can be seen, similarly to the survival case, the transition probability coefficients  of $P_{e\mu}=|A_{\mu e}|^2$ can be obtained
from those of $P_{\mu e}$ through the transformations
$u_{12}\leftrightarrow u_{21}$ and $\phi\to-\phi$.
\end{itemize}

\paragraph{Automatic Boundedness for $\Delta \neq 0$.}
The long-time behaviour of the transition probabilities is controlled
by the factor
\[
e^{-\aabs t}|\sin(\Lambda t)|^2 \sim \frac{1}{4}e^{(2|k|-\aabs)t},
\]
where $\Lambda = w + ik$, so boundedness requires \(2|k|\le \aabs\).
Using
\[
|\Lambda^2| = \sqrt{Q^2+\aabs^2\Delta^2},
\qquad
Q \equiv w^2-k^2=\Delta^2+\lambda^2-\frac{\aabs^2}{4},
\]
we obtain
\[
k^2=\frac{\sqrt{Q^2+\aabs^2\Delta^2}-Q}{2}.
\]
Substituting into \(4k^2\le \aabs^2\) yields
\[
2\sqrt{Q^2+\aabs^2\Delta^2}\le 2Q+\aabs^2.
\]
Since the right-hand side is non-negative, squaring gives
\[
0\le 4\aabs^2\lambda^2,
\]
which is identically satisfied within the physical parameter domain considered. Therefore, for \(\Delta\neq 0\), the evolution is automatically bounded for any real $\lambda$, provided $\aabs \ge 0$. No additional stability condition arises in this regime.

In this work, we do not pursue the full analytical expressions of the oscillation probabilities, as our focus is on the behavior of the $CP$ and $CPT$ asymmetries, which will be discussed in the following sections.

\subsection{Phenomenological regimes}
\label{sec:regimes}
Our framework encompasses several distinct physical regimes depending on the relative magnitudes of $\Delta$, $\lambda$, and $\aabs$. In this section, we explore these regimes by analyzing the eigenvalues of the Hamiltonian, Eq.~(\ref{Enereigenvalues}).

\subsubsection{Pedagogical case}
\textbf{Condition: $\Delta=0$, $\delta=0$.}

To build intuition, this simplified case is considered, in which the eigenvalues of Eq.~\eqref{Enereigenvalues} reduce to
\begin{equation}
E_\pm = -\frac{i\aabs}{2} \pm \sqrt{\lambda^2 - \frac{\aabs^2}{4}},
\end{equation}
and consequently $\Lambda = \sqrt{\lambda^2 - \aabs^2/4}$. 

\textbf{Physical interpretation:}
From the latter expression of $\Lambda$, it is clear that the relation between $2 \lambda$ and $\aabs$ governs the \emph{qualitative behavior} of the system.

\begin{itemize}
\item \textbf{Underdamped} ($2\lambda>\aabs$): $\Lambda\in\mathbb{R}$,
    dynamics oscillatory. Genuine quantum flavour oscillations are sustained.
\item \textbf{Critical Point} ($2\lambda=\aabs$): $\Lambda=0$,
    polynomial--times--exponential behaviour.
\item \textbf{Overdamped} ($2\lambda<\aabs$): $\Lambda$ purely imaginary
    at $\Delta=0$, no oscillatory behaviour; the system decays
    monotonically without flavour oscillations.
\end{itemize}
The analogy with the damped harmonic oscillator is exact. In
particular, at the critical point the flavour oscillation
probabilities are
\begin{align}
P_{\nu_e\to\nu_e}(t) &= e^{-\aabs t}\left[1+\aabs t\cos 2\theta
  +\frac{\aabs^2 t^2}{4}\left(1-\sin^2 2\theta\,\sin^2\phi\right)\right],\\
P_{\nu_\mu\to\nu_\mu}(t) &= e^{-\aabs t}\left[1-\aabs t\cos 2\theta
  +\frac{\aabs^2 t^2}{4}\left(1-\sin^2 2\theta\,\sin^2\phi\right)\right],\\
P_{\nu_e\to\nu_\mu}(t) &= e^{-\aabs t}\,\frac{\aabs^2 t^2}{4}
  \left(1+\sin 2\theta\,\sin\phi\right)^2,\\
P_{\nu_\mu\to\nu_e}(t) &= e^{-\aabs t}\,\frac{\aabs^2 t^2}{4}
  \left(1-\sin 2\theta\,\sin\phi\right)^2.
\end{align}
For $\theta=0$ these reduce to the mass-basis results
$|u_{11}|^2=e^{-\aabs t}(1+\aabs t/2)^2$,
$|u_{22}|^2=e^{-\aabs t}(1-\aabs t/2)^2$ and
$|u_{12}|^2=|u_{21}|^2=e^{-\aabs t}(\aabs t/2)^2$.
The forward and backward transition probabilities differ,
\begin{equation}
P_{\nu_e\to\nu_\mu}-P_{\nu_\mu\to\nu_e}
= \aabs^2 t^2\, e^{-\aabs t}\sin 2\theta\,\sin\phi ,
\end{equation}
in agreement with the $\Lambda\to 0$ limit of the T asymmetry,
Eq.~\eqref{eq:DeltaT}, at $\delta=0$.
Notice that the
usual oscillatory $t$-dependence is replaced by exponential decay
modulated by polynomials in $t$---the characteristic signature of
this configuration.

\subsubsection{Quasi-global damping (neutrino-loss scenario)}
\label{nuloss}

\textbf{Condition:} $\aabs\ll\Omega$, where
$\Omega=\sqrt{\Delta^2+\lambda^2}$.

In this regime terms of order $\aabs^2$ in the expression for
$\Lambda^2$ can be neglected, giving
\begin{equation}
\Lambda \simeq \sqrt{\Omega^2 + i\,\aabs\Delta}.
\label{eq:Lambda2_general}
\end{equation}
Expanding to first order in $\aabs\Delta/\Omega^2\ll 1$:
\begin{equation}
\Lambda \simeq \Omega + i\,\frac{\aabs\Delta}{2\Omega}.
\label{eq:Lambda_approx}
\end{equation}
Substituting into Eq.~\eqref{Enereigenvalues}:
\begin{equation}
E_\pm \simeq \pm\Omega - \frac{i\aabs}{2}
    \!\left(1\mp\frac{\Delta}{\Omega}\right).
\label{eq:Epm_general}
\end{equation}
In the limit $|\Delta|\ll\Omega$ the correction proportional to
$\Delta/\Omega$ is subleading; neglecting it:
\begin{equation}
E_\pm \simeq \pm\Omega - \frac{i\aabs}{2}.
\label{eq:Epm_global_damping}
\end{equation}
\textbf{Physical interpretation:} This is the global damping regime,
where $\Gamma_{\rm eff}\to \aabs\mathbb{I}$, which renders the Hamiltonian
normal. The oscillation probability factorises as
\begin{equation}
P_{\nu_\alpha\to\nu_\beta}
= e^{-\aabs L}\,P^{\mathrm{std}}_{\nu_\alpha\to\nu_\beta},
\end{equation}
recovering the neutrino-loss scenario of Ref.~\cite{neutrino_loss_ref}.

\subsubsection{Selective damping (standard invisible neutrino decay)}

\textbf{Condition:} $\lambda\ll\sqrt{\Delta^2+\aabs^2/4}$.

For small $\lambda$,
\begin{equation}
\Lambda^2 = \Delta^2 + i\Delta\aabs - \frac{\aabs^2}{4}
\approx \left(\Delta + i\frac{\aabs}{2}\right)^2,
\qquad\text{so}\qquad
\Lambda \approx \Delta + i\frac{\aabs}{2},
\end{equation}
and the eigenvalues become:
\begin{equation}
E_+ = \Delta, \qquad E_- = -\Delta - i\aabs.
\end{equation}
\textbf{Physical interpretation:} Only the heavier mass eigenstate
$\nu_2$ decays, while the lighter one remains stable, reproducing
the standard invisible-decay scenario in which $\nu_2$ decays into
states that are not observed. Since the model contains no
regeneration terms, no lighter active neutrino is produced in the
decay.
The absorptive parameter is related to the lifetime $\tau_2$ of
$\nu_2$ in its rest frame by
\begin{equation}
\aabs \equiv \frac{m_2}{2\,\tau_2 E_\nu},
\end{equation}
so that $2\aabs = m_2/(\tau_2 E_\nu)$ is the lab-frame decay rate
of $\nu_2$.

\section{$CP$, $T$ and $CPT$ Asymmetries}
\label{sec:asymmetries}

We define:
\begin{align}
\Delta_{\rm CP}^{\alpha\beta}
&=
P_{\nu_\alpha\rightarrow\nu_\beta}
-
P_{\bar{\nu}_\alpha\rightarrow\bar{\nu}_\beta},
\\[2mm]
\Delta_{\rm T}^{\alpha\beta}
&=
P_{\nu_\alpha\rightarrow\nu_\beta}
-
P_{\nu_\beta\rightarrow\nu_\alpha},
\\[2mm]
\Delta_{\rm CPT}^{\alpha\beta}
&=
P_{\nu_\alpha\rightarrow\nu_\beta}
-
P_{\bar{\nu}_\beta\rightarrow\bar{\nu}_\alpha}.
\end{align}
where for the survival we have: 
\begin{equation}
 \Delta_{\rm CP}^{\alpha\alpha}
=
\Delta_{\rm CPT}^{\alpha\alpha}   
\end{equation}
while for flavour transition $(\alpha \ne \beta)$:
\begin{align}
\Delta_{\rm CP}^{\alpha\beta}
&\neq
\Delta_{\rm CPT}^{\alpha\beta}.
\end{align}

 In the following we adopt the prescription for the complex phase conjugation.

\subsection{Fixed background prescription: $\phi\to-\phi$ and $\delta \to \delta$}
The two phases $\phi$ and $\delta$ involved in our system have
qualitatively different origins, which determines how they transform
under neutrino--antineutrino exchange. The phase $\phi$ belongs to
the flavour rotation matrix $R(\theta,\phi)$: it encodes the quantum
interference between flavour eigenstates in the active-neutrino sector
and changes sign under the standard complex conjugation
$U\to U^*$ that maps $\nu\to\bar\nu$. The phase $\delta$, by contrast,
is an intrinsic property of the coherent background field: it
parametrises the orientation of the field configuration in the complex
plane of its coupling to neutrinos. Once a particular background
configuration is selected, this orientation is fixed independently of
the particle sector. Neutrinos and antineutrinos therefore propagate
in the same physical background, so that exchanging $\nu$ and
$\bar\nu$ conjugates the particle sector without conjugating the
background configuration itself.

A possible microscopic realization of this picture arises if the
coherent background originates from a field acquiring a complex vacuum
expectation value,
\[
    \langle \Phi \rangle = v e^{i\delta}.
\]
Once the vacuum selects a particular configuration, the phase $\delta$
is frozen by this vacuum choice and acts as a fixed background parameter
for neutrino propagation. It is important to distinguish the
neutrino--antineutrino exchange considered here, in which the same
vacuum configuration is held fixed, from a $CP$ or $CPT$ transformation
of the complete system. In the latter case, the background itself would
also have to be transformed according to its microscopic transformation
properties; for a complex scalar vacuum expectation value, for example,
$CP$ would map $\langle\Phi\rangle=v e^{i\delta}$ into
$\langle\Phi\rangle^*=v e^{-i\delta}$.

This viewpoint is analogous to the treatment of fixed Lorentz- and
$CPT$-violating background coefficients in the Standard Model Extension
\cite{Colladay:1998fh,Kostelecky:2003cr}, where background tensor
coefficients may originate from vacuum expectation values that select
preferred directions while remaining fixed under transformations of
the particle sector. In our case, $\delta$ plays the role of such a
background orientation. More generally, if the coherent field arises
from a misalignment between the interaction eigenbasis of the background
and the neutrino mass basis, as may occur for axion-like fields
\cite{Huang:2018cwo}, $CPT$-odd interaction sectors
\cite{Barger:2000iv}, or equivalence-principle-violating potentials
\cite{Diaz:2020aax}, then $\delta$ parametrises this misalignment in
the complex plane. Different physical configurations of the background
may correspond to different values of $\delta$, but conjugating the
neutrino sector does not rotate the selected background configuration.

The prescription
\[
    \phi\to-\phi,\qquad \delta\to\delta
\]
is therefore physically motivated: it follows from comparing
neutrinos and antineutrinos propagating in the same fixed background.
The phase $\phi$, being associated with the particle mixing sector, is
complex conjugated, whereas $\delta$, being associated with the selected
background configuration, remains fixed.   As shown below, this differential 
transformation is the origin of the
  $CPT$ asymmetries generated by this mechanism: if $\delta$ were also
  reversed, they would vanish identically.

Thus, under the above prescription we have:

\begin{equation}
   \Delta_{\rm CP}^{\alpha\beta}=
2A^{\alpha\beta}_{s}\sin\phi + 2A^{\alpha\beta}_{2s}\sin 2\phi
\end{equation}
Where the list of asymmetries is:
\begin{equation}
\Delta_{\rm CP}^{e\mu}
=
2\lambda \sin 2\theta \sin\phi\,
\frac{e^{-\aabs t}}{|\Lambda|^2}
\left|\sin(\Lambda t)\right|^2
\left[
2C_{+}\sin\delta
+
\aabs\cos\delta
\right].
\end{equation}
\begin{equation}
\Delta_{\rm CP}^{\mu e}
=
2\lambda \sin 2\theta \sin\phi\,
\frac{e^{-\aabs t}}{|\Lambda|^2}
\left|\sin(\Lambda t)\right|^2
\left[
2C_{+}\sin\delta
-
\aabs\cos\delta
\right].
\end{equation}
\begin{equation}
\Delta_{\rm CPT}^{ee}
=
-2\lambda \sin 2\theta \sin\phi \sin\delta\,
\frac{e^{-\aabs t}}{|\Lambda|^2}
\left[
2\left|\sin(\Lambda t)\right|^2 C_{+}
-
A(t)
\right].
\end{equation}
\begin{equation}
 \Delta_{\rm CP}^{ee}
=\Delta_{\rm CPT}^{ee}   
\end{equation}
\begin{equation}
\Delta_{\rm CPT}^{\mu\mu}
=
-2\lambda \sin 2\theta \sin\phi \sin\delta\,
\frac{e^{-\aabs t}}{|\Lambda|^2}
\left[
2\left|\sin(\Lambda t)\right|^2 C_{+}
+
A(t)
\right].
\end{equation}
\begin{equation}
 \Delta_{\rm CP}^{\mu\mu}
=\Delta_{\rm CPT}^{\mu\mu}   
\end{equation}
\begin{equation}
\Delta_{\rm CPT}^{e\mu}
=
2\lambda \sin 2\theta \sin\delta\,
\frac{e^{-\aabs t}}{|\Lambda|^2}
\left|\sin(\Lambda t)\right|^2
\left[
2C_{+}\sin\phi
+
\aabs\cos\phi
\right].
\end{equation}
\begin{equation}
\Delta_{\rm CPT}^{\mu e}
=
2\lambda \sin 2\theta \sin\delta\,
\frac{e^{-\aabs t}}{|\Lambda|^2}
\left|\sin(\Lambda t)\right|^2
\left[
2C_{+}\sin\phi
-
\aabs\cos\phi
\right].
\end{equation}

\begin{equation}
\Delta_{\rm T}^{e\mu}
=
2\aabs \lambda \sin 2\theta\,
\frac{e^{-\aabs t}}{|\Lambda|^2}
\left|\sin(\Lambda t)\right|^2
\sin(\delta+\phi).
\end{equation}
\begin{equation}
\Delta_{\rm T}^{\mu e}
=
-\Delta_{\rm T}^{e\mu}.
\end{equation}

where
\begin{align}
A(t) &= k\sin(2\omega t) - \omega\sinh(2kt),\\
C_{+}
&=
\Delta \cos 2\theta
+
\lambda \sin 2\theta \cos\delta \cos\phi .
\end{align}

\begin{figure}[t]
  \centering
  \begin{subfigure}{0.48\linewidth}
    \centering
    \includegraphics[width=\linewidth]{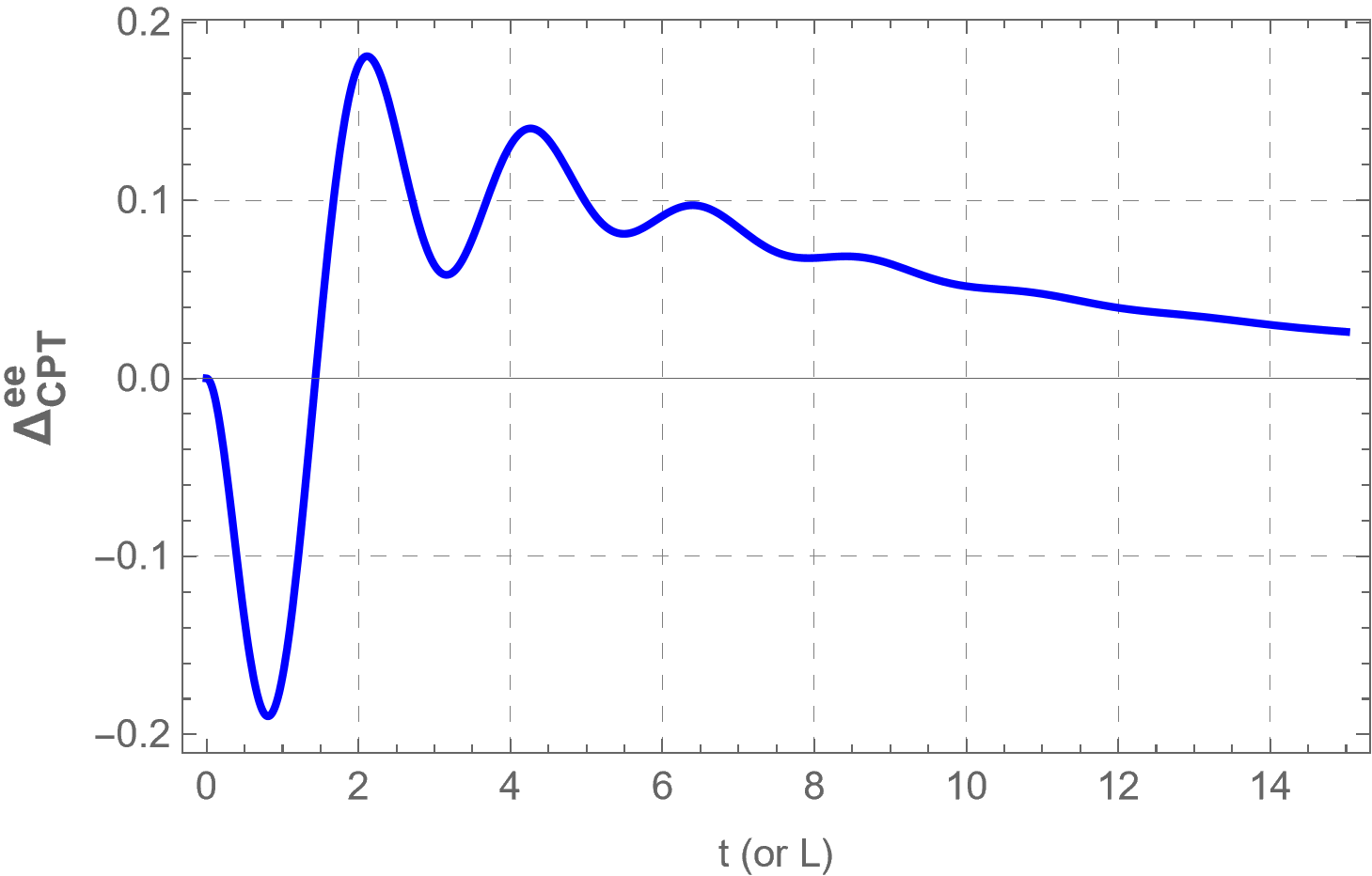}
  \end{subfigure}
  \hfill
  \begin{subfigure}{0.48\linewidth}
    \centering
    \includegraphics[width=\linewidth]{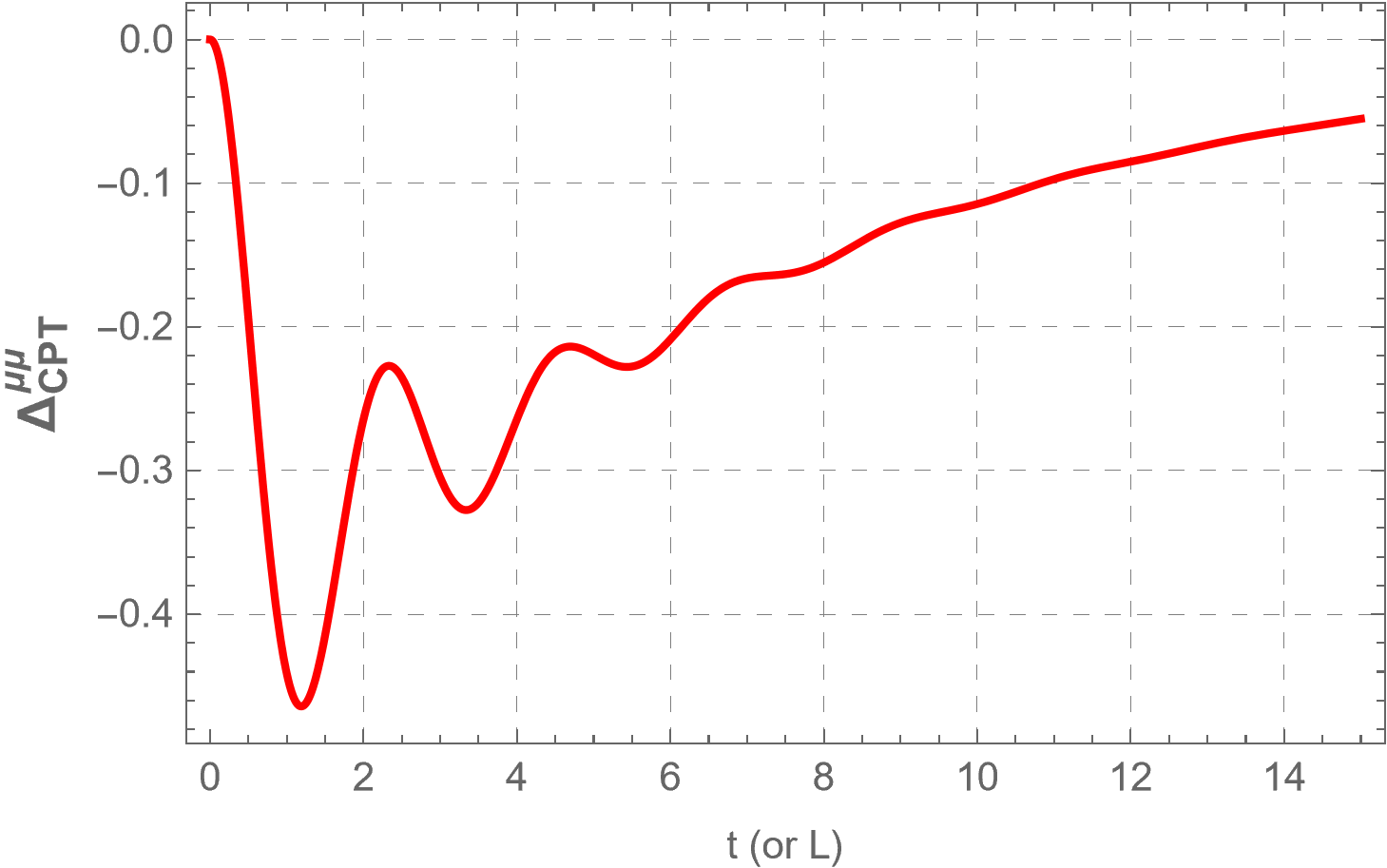}
  \end{subfigure}

  \vspace{0.4cm}

  \begin{subfigure}{0.48\linewidth}
    \centering
    \includegraphics[width=\linewidth]{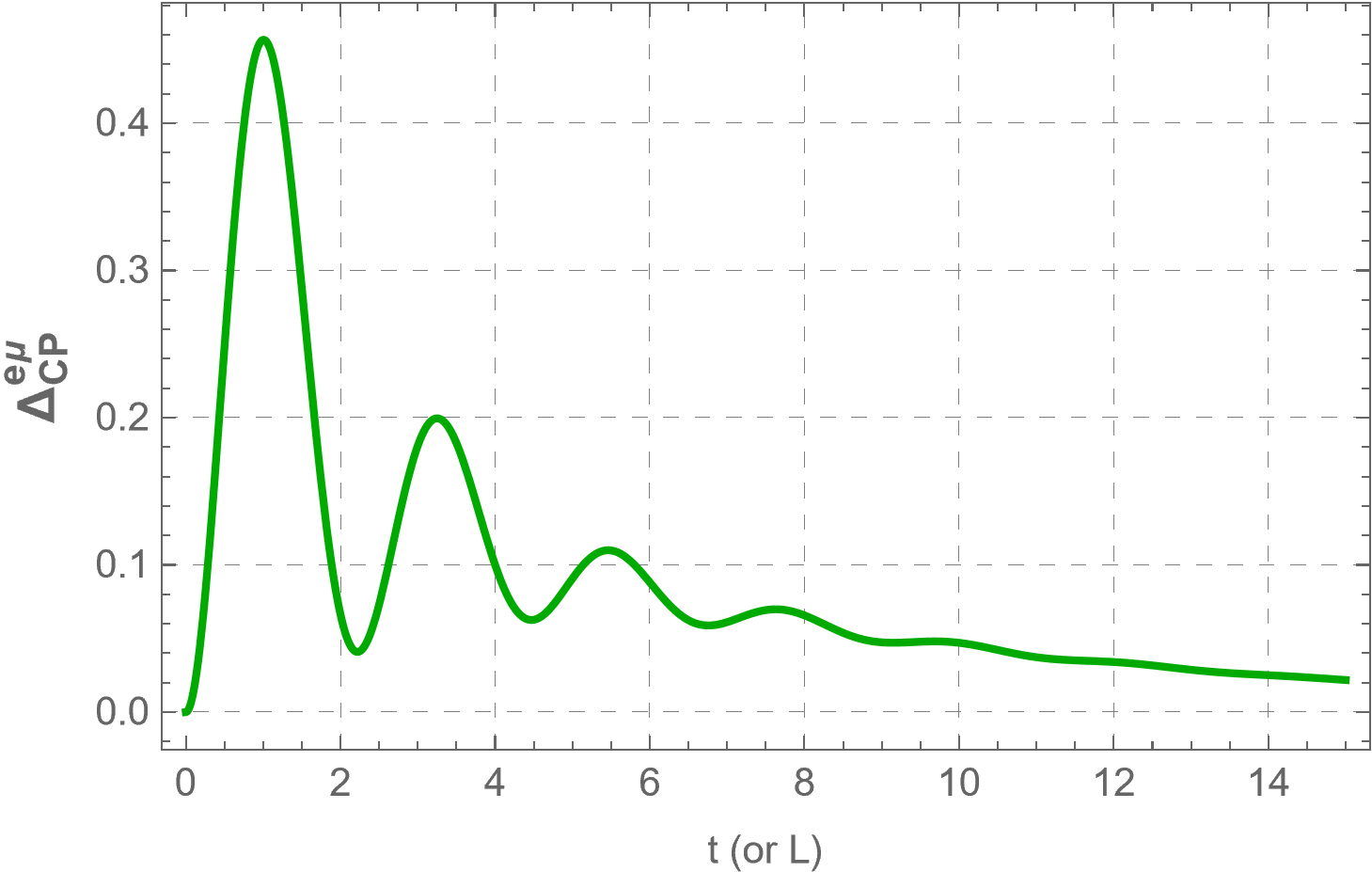}
  \end{subfigure}
  \hfill
  \begin{subfigure}{0.48\linewidth}
    \centering
    \includegraphics[width=\linewidth]{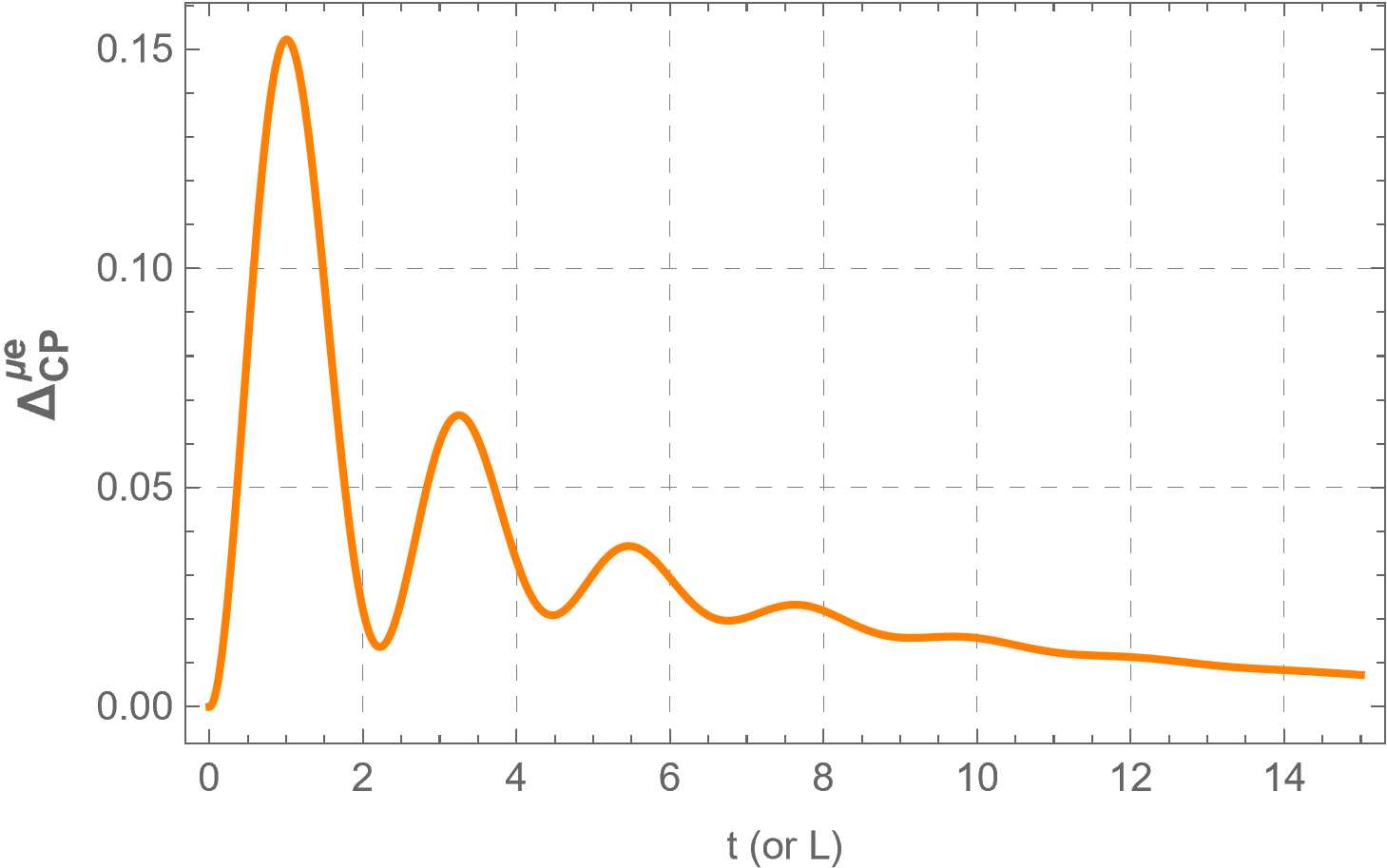}
  \end{subfigure}
\caption{Time evolution of the $CPT$ asymmetries
$\Delta_{\rm CPT}^{ee}$ (top-left) and $\Delta_{\rm CPT}^{\mu\mu}$
(top-right), and of the $CP$ asymmetries $\Delta_{\rm CP}^{e\mu}$
(bottom-left) and $\Delta_{\rm CP}^{\mu e}$ (bottom-right), for
$\theta=\pi/4$, $\Delta=-1$, $\lambda=1$, $\aabs=0.5$, and
$\delta=\phi=\pi/4$. The curves illustrate the interplay between
oscillatory behaviour and damping induced by the non-Hermitian
evolution.}
  \label{fig:asymm_time}
\end{figure}

\begin{figure}[t]
  \centering
  \begin{subfigure}{0.48\linewidth}
    \centering
    \includegraphics[width=\linewidth]{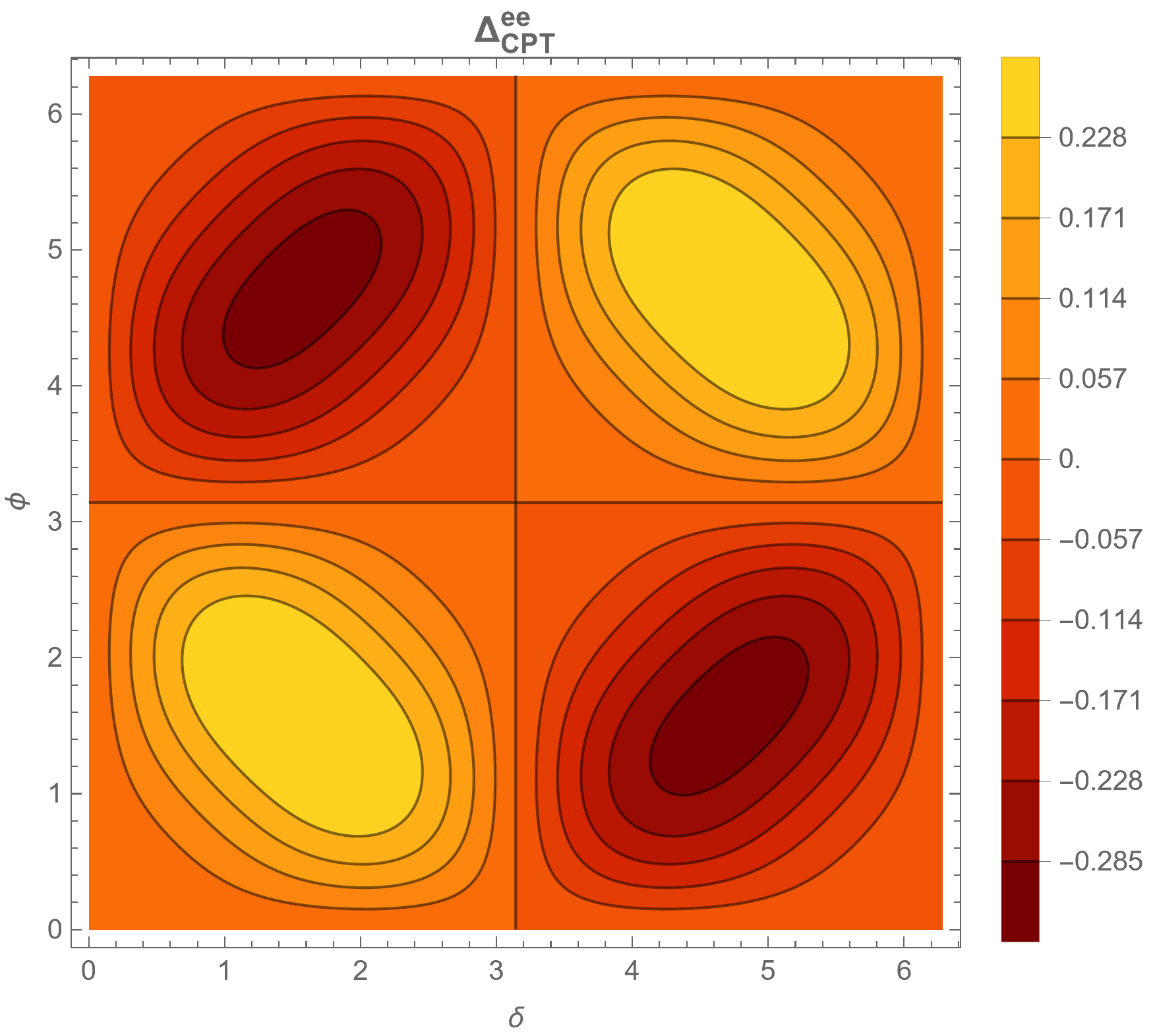}
  \end{subfigure}
  \hfill
  \begin{subfigure}{0.48\linewidth}
    \centering
    \includegraphics[width=\linewidth]{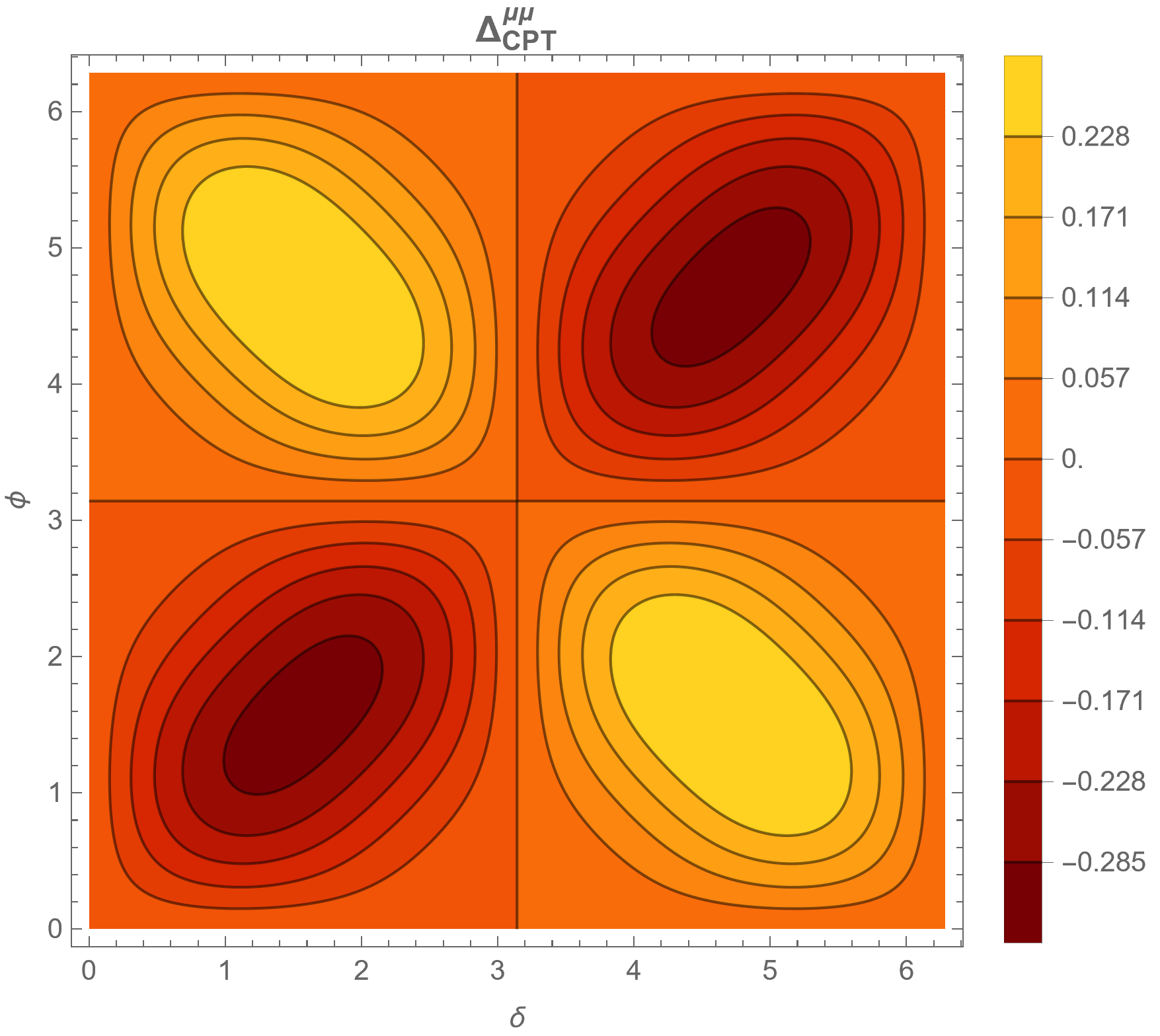}
  \end{subfigure}

  \vspace{0.4cm}

  \begin{subfigure}{0.48\linewidth}
    \centering
    \includegraphics[width=\linewidth]{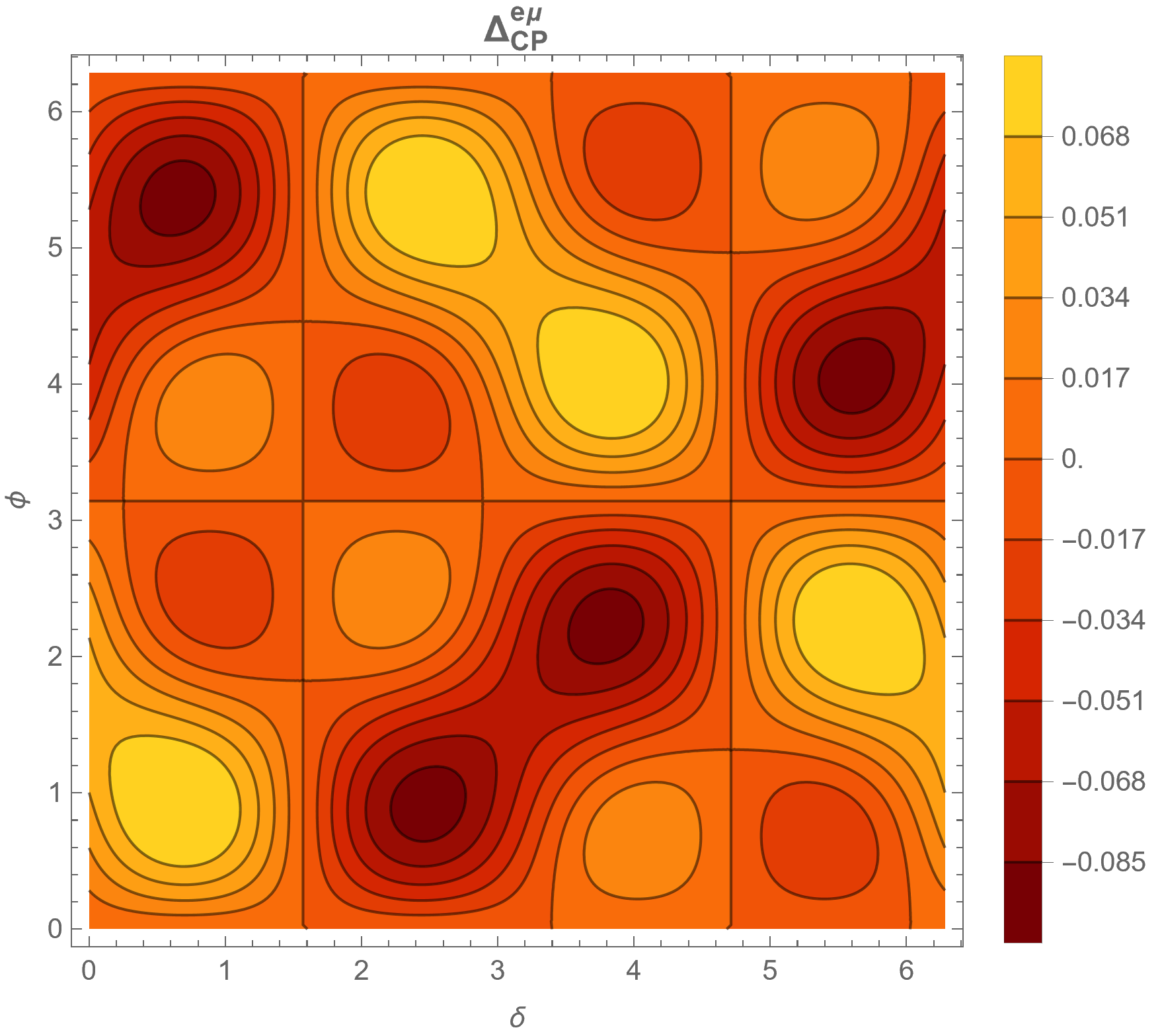}
  \end{subfigure}
  \hfill
  \begin{subfigure}{0.48\linewidth}
    \centering
    \includegraphics[width=\linewidth]{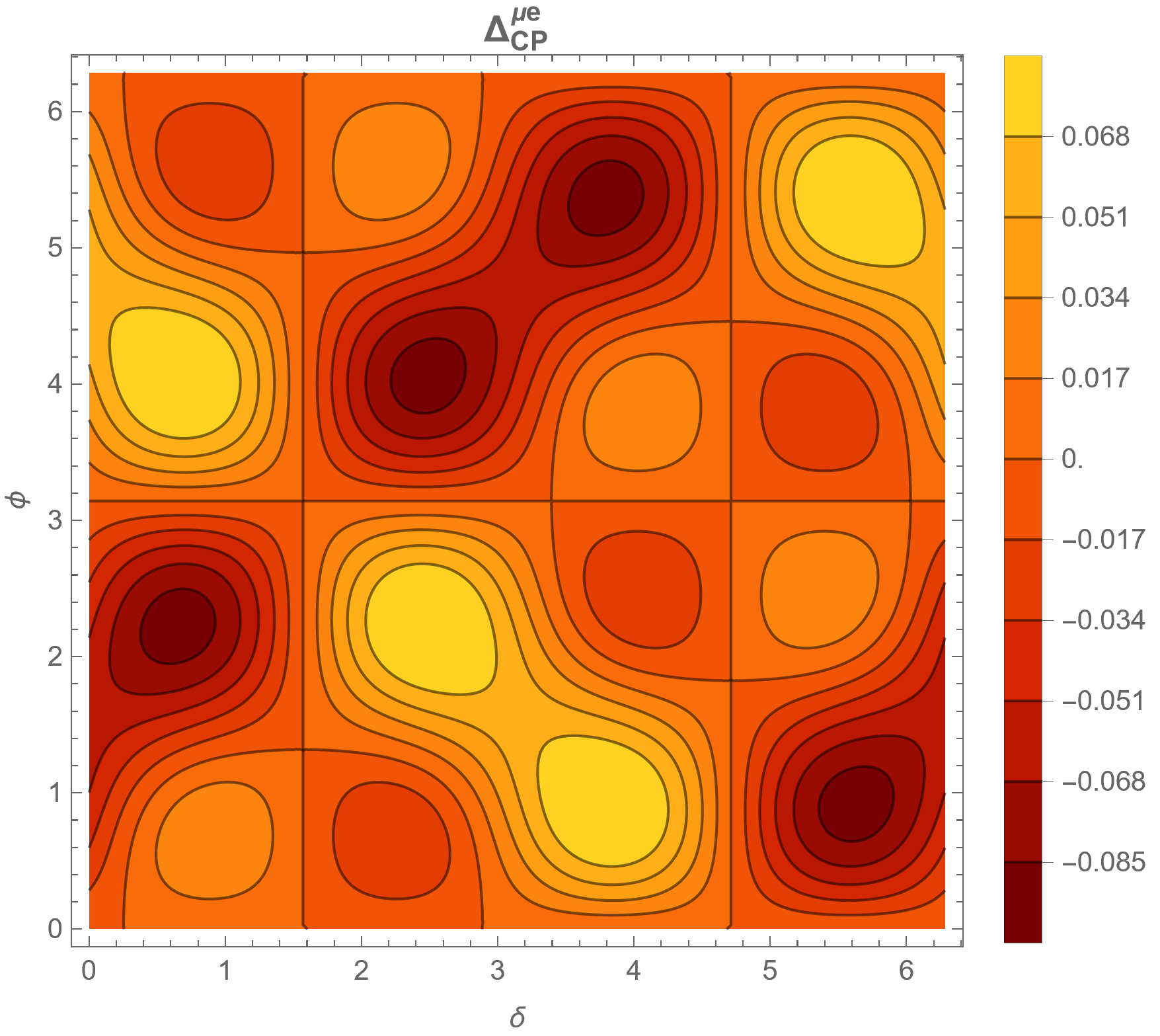}
  \end{subfigure}
\caption{Contour plots of $\Delta_{\rm CPT}^{ee}$ (top-left),
$\Delta_{\rm CPT}^{\mu\mu}$ (top-right), $\Delta_{\rm CP}^{e\mu}$
(bottom-left), and $\Delta_{\rm CP}^{\mu e}$ (bottom-right) in the
$(\delta,\phi)$ plane at $t=5$, for $\theta=\pi/4$, $\Delta=-1$,
$\lambda=0.1$, $\aabs=0.5$.}
  \label{fig:contours}
\end{figure}

\begin{figure}[htb]
  \centering
  \includegraphics[width=0.5\linewidth]{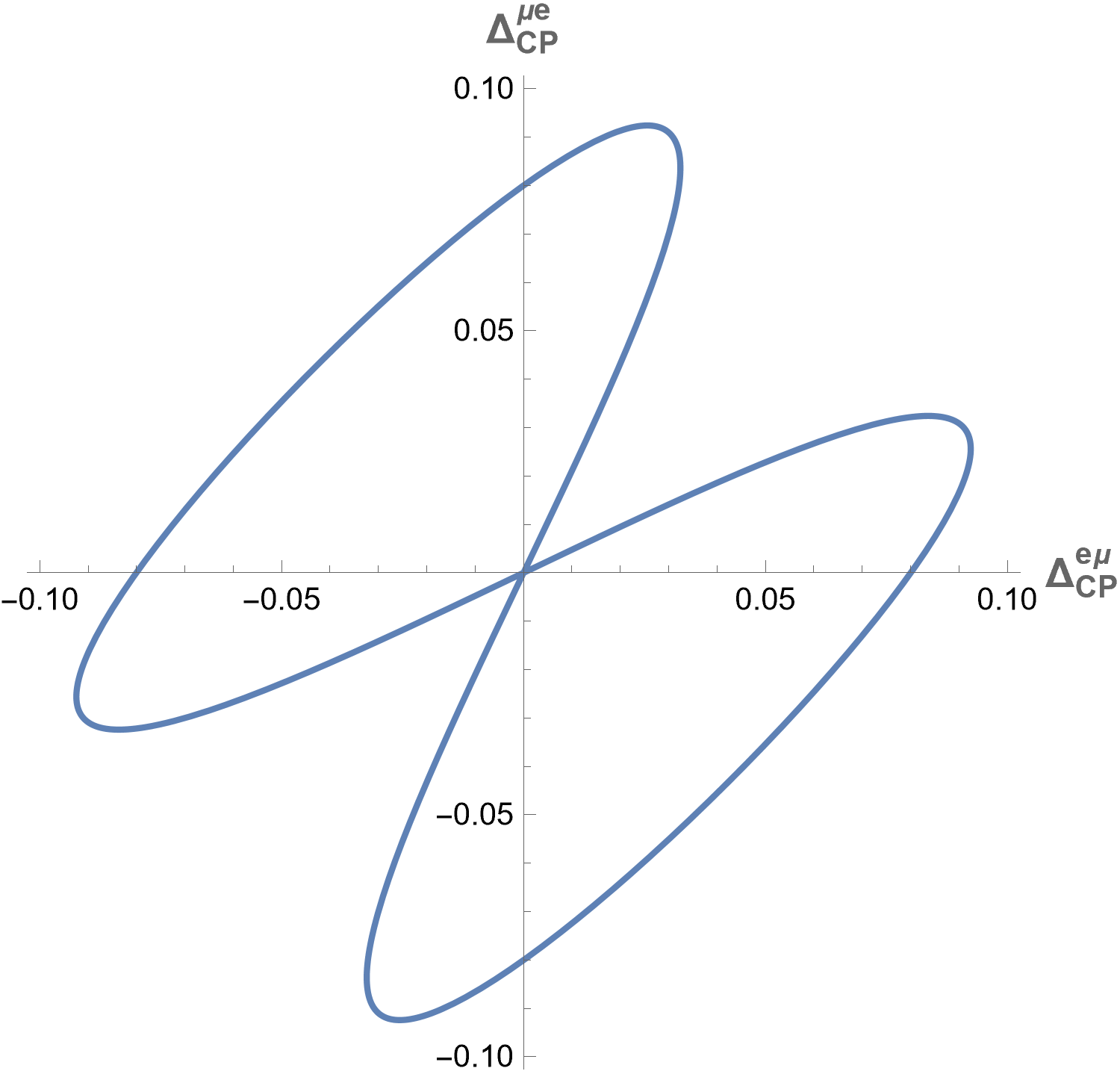}
\caption{Parametric curve
$(\Delta_{\rm CP}^{e\mu},\Delta_{\rm CP}^{\mu e})$ as $\phi$ varies
from $0$ to $2\pi$ at $t=5$, $\delta=\pi/4$. Same parameters as
Fig.~\ref{fig:contours}.}
  \label{fig:lissajous}
\end{figure}

\subsection{CP and CPT asymmetry}
A simple inspection of all the asymmetries, including the T asymmetry, indicates that to avoid vanishing, 
all of them require that the off diagonal background field $\lambda\neq 0$ and $\theta\neq n\pi/2$, where $n$ is an integer. Another characteristic shared by all asymmetries is 
the damping factor $e^{-\aabs t}$ partially offset by the growth of $|\sin(\Lambda t)|^2$.
\subsubsection{CP asymmetry}
Aside from the general features mentioned earlier, all the CP asymmetries need that 
$\phi\neq n\pi$ to be non-vanishing. The following special cases are instructive:
\begin{itemize}
  \item $\theta=\pi/4$, $\delta=(2n+1)\pi/2$:
      $\Delta_{\rm CP}^{e\mu}=\Delta_{\rm CP}^{\mu e}=0$ and
      $\Delta_{\rm CP}^{ee}=-\Delta_{\rm CP}^{\mu\mu}$.
\item $\delta=n\pi$:
$\Delta_{\rm CP}^{ee }=\Delta_{\rm CP}^{\mu\mu }=0$ and
$\Delta_{\rm CP}^{e\mu }
    =-\Delta_{\rm CP}^{\mu e}$.
\item $\delta=\pm\pi/2$:
     $\Delta_{\rm CP}^{e\mu }
    =\Delta_{\rm CP}^{\mu e}$.
\end{itemize}
\subsubsection{CPT asymmetry}
All the CPT asymmetries require that 
$\delta\neq n\pi$ for being non-null. In that sense, the following cases are also worth pointing out:
 \begin{itemize}
\item $\theta=\pi/4$, $\delta=(2n+1)\pi/2$:
    $\Delta_{\rm CPT}^{e\mu}=-\Delta_{\rm CPT}^{\mu e}$ and
    $\Delta_{\rm CPT}^{ee}=-\Delta_{\rm CPT}^{\mu\mu}$; if in addition
    $\aabs=0$ or $\phi=(2n+1)\pi/2$, then
    $\Delta_{\rm CPT}^{e\mu}=\Delta_{\rm CPT}^{\mu e}=0$.
\item $\phi=n\pi$:
    $\Delta_{\rm CPT}^{e\mu}=-\Delta_{\rm CPT}^{\mu e}$ and
    $\Delta_{\rm CPT}^{ee}=\Delta_{\rm CPT}^{\mu\mu}=0$.
\item $\phi=(2n+1)\pi/2$:
    $\Delta_{\rm CPT}^{e\mu}=\Delta_{\rm CPT}^{\mu e}$.
\end{itemize}
Interestingly, for $\theta=\pi/4$ and $\delta=(2n+1)\pi/2$,
$\Delta_{\rm CPT}^{e\mu(\mu e)}$ is controlled by the absorptive
parameter $\aabs$.

Figure~\ref{fig:asymm_time} shows the time evolution of four 
asymmetries ($\Delta_{\rm CPT}^{ee}$, $\Delta_{\rm CPT}^{\mu\mu}$, $\Delta_{\rm CP}^{e \mu}$ and $\Delta_{\rm CP}^{\mu e}$)  for representative parameter values, illustrating the
interplay between oscillatory and exponentially damped contributions.

\subsubsection{Parameter-space analysis}
In the parameter-space analysis of the asymmetries in terms of
$\delta$ and $\phi$, we assume the relation $\lambda=0.1|\Delta|$.
This choice reflects a regime where the pumping effects are
subdominant relative to standard oscillations yet large enough
to produce observable asymmetries, representing a phenomenologically
conservative. 
Figure~\ref{fig:contours} shows the $CP$ and $CPT$ asymmetries in the
$\delta$--$\phi$ plane, evaluated at
$\theta = \pi/4, \Delta = -1$ and $\aabs = 0.5$, with $\delta$ and $\phi$ 
varying over
$[0,2\pi]$.
For these parameter values, with $\lambda=0.1|\Delta|$
and $\theta=\pi/4$, we have
$C_+=\lambda\cos\delta\cos\phi$.
Consequently, the contributions proportional to
$\lambda C_+$ are of order $\lambda^2$ and are
subleading. To leading order in $\lambda/|\Delta|$,
the CPT asymmetries therefore satisfy
$\Delta_{\rm CPT}^{ee}\simeq
-\Delta_{\rm CPT}^{\mu\mu}$
and are proportional to $\sin\phi\sin\delta$. This functional dependence explains
the symmetry pattern observed in the contours: the oval shapes
in the first--second $\delta$--$\phi$ quadrants are the same as
those in the third--fourth quadrants.  The same argument applies to mixed-sign regions: the oval contours in the
first--second $\delta$ quadrants and third--fourth $\phi$ quadrants are
identical to those in the opposite $\delta$--$\phi$ quadrants. Moreover, the top--bottom interchange of the oval shapes between
$\Delta_{\rm CPT}^{ee}$ and $\Delta_{\rm CPT}^{\mu\mu}$ is expected, given the opposite
sign relation between these asymmetries. 

Figure~\ref{fig:contours} also shows the $CP$ asymmetries
 $\Delta_{\rm CP}^{e \mu}$ and $\Delta_{\rm CP}^{\mu e}$. For the chosen set of 
parameters, these asymmetries satisfy 
$\Delta_{\rm CP}^{e \mu} \propto
\sin\phi\,\cos\delta\left(\aabs +2\lambda\sin\delta\cos\phi\right)$,
whereas
$\Delta_{\rm CP}^{\mu e} \propto
-\sin\phi\,\cos\delta\left(\aabs -2\lambda\sin\delta\cos\phi\right)$.
From these expressions, the relation between the contour patterns of both
asymmetries can be directly inferred. In particular, under the shift
$\delta \to \delta+\pi$, one finds
$\Delta_{\rm CP}^{e \mu}(\delta,\phi)=\Delta_{\rm CP}^{\mu e}(\delta+\pi,\phi)$,
up to the same overall proportionality factor. Therefore, for the whole
range of $\phi$, the structure of $\Delta_{\rm CP}^{e \mu}$ when $\delta$ lies in
the first (second) quadrant matches the structure of $\Delta_{\rm CP}^{\mu e}$
when $\delta$ lies in the third (fourth) quadrant, and vice versa.

The non-trivial interplay between the two $CP$ asymmetries is
illustrated in Fig.~\ref{fig:lissajous}, which shows the parametric
curve $(\Delta_{\rm CP}^{e \mu},\Delta_{\rm CP}^{\mu e})$ as $\phi$ varies over
$[0,2\pi]$. As expected, the parametric curve contains the coordinate pair $(0,0)$, while a general tendency of $\Delta_{\rm CP}^{e \mu}\sim -\Delta_{\rm CP}^{\mu e}$ is observed. 

All the observations made above for the plots for  $\Delta_{\rm CP}^{e \mu}$ and $\Delta_{\rm CP}^{\mu e}$ 
can be directly applied for the case of 
$\Delta_{\rm CPT}^{e \mu}$ and $\Delta_{\rm CPT}^{\mu e}$ 
by only exchanging $\delta \leftrightarrow \phi$.

\subsection{T asymmetry}

\begin{figure}[t]
  \centering
  \includegraphics[width=0.65\linewidth]{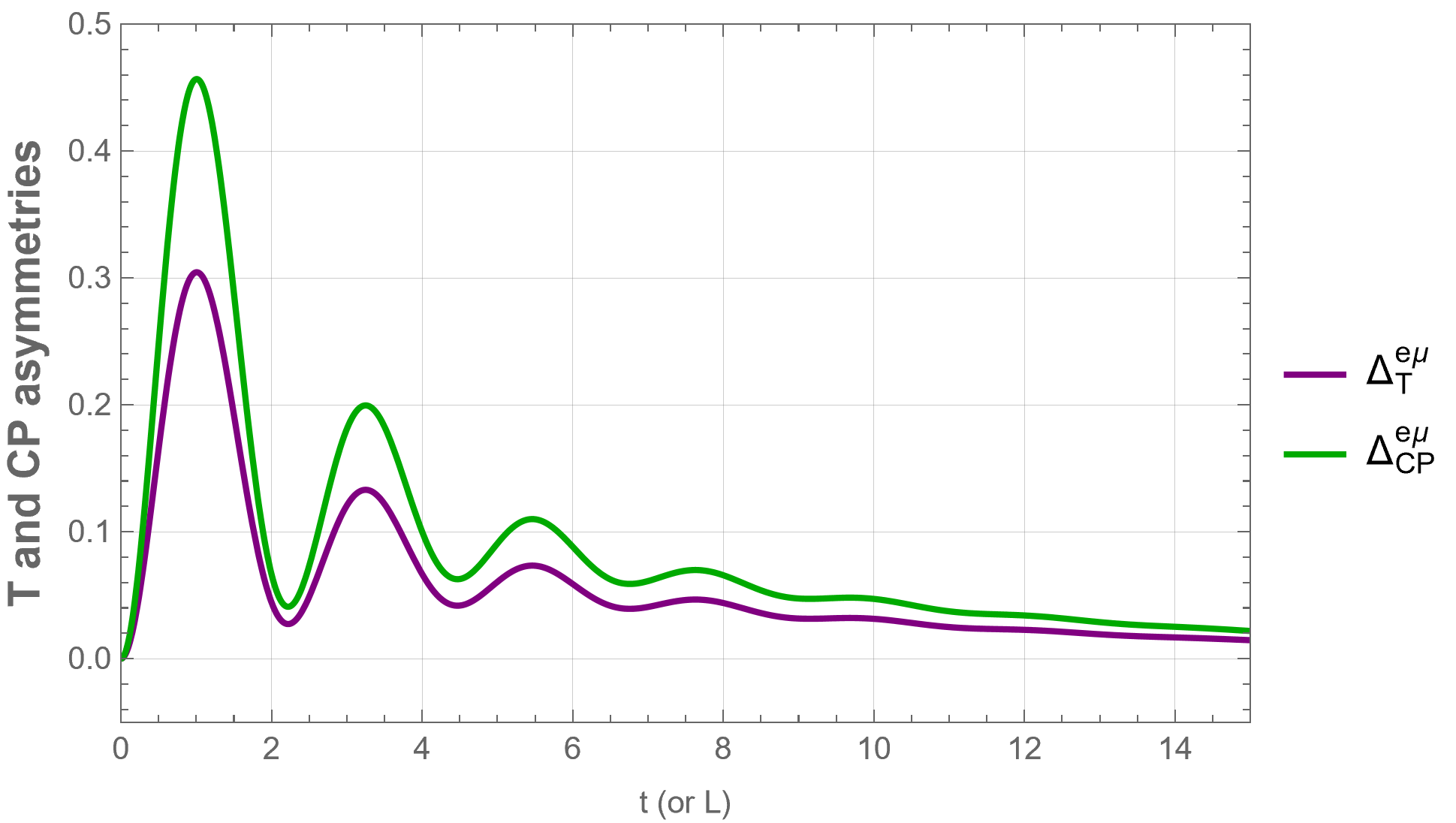}
\caption{Time evolution of the T asymmetry $\Delta_{\rm T}^{e\mu}$
(purple) and the CP asymmetry $\Delta_{\rm CP}^{e\mu}$ (green), for
the same parameters as Fig.~\ref{fig:asymm_time}. Both carry the
same oscillatory and damping structure, but differ in amplitude:
$\Delta_{\rm T}^{e\mu}$ is proportional to
$\aabs\sin(\delta+\phi)$ and is positive throughout for these
parameter values, while $\Delta_{\rm CP}^{e\mu}$ receives additional
contributions proportional to $\sin\delta$ and $\cos\delta$. The
T asymmetry isolates the absorptive coupling in a single
observable.}
  \label{fig:DeltaT}
\end{figure}

The framework also generates a non-trivial T asymmetry, defined as
the difference between the forward and backward transition
probabilities:
\begin{equation}
\Delta_{\rm T}^{e\mu}
\equiv P_{\nu_e\to\nu_\mu} - P_{\nu_\mu\to\nu_e}
= 2\aabs\lambda\sin 2\theta\,
\frac{e^{-\aabs t}|\sin(\Lambda t)|^2}{|\Lambda|^2}
\sin(\delta+\phi),
\qquad
\Delta_{\rm T}^{\mu e} = -\Delta_{\rm T}^{e\mu}.
\label{eq:DeltaT}
\end{equation}
Several properties distinguish this asymmetry from the $CP$ and
$CPT$ ones:
\begin{itemize}
\item It requires $\aabs\neq 0$ explicitly --- the absorptive
    channel is essential, and $\Delta_{\rm T}^{e\mu}$ is directly
    proportional to $\aabs$. It therefore provides a clean
    experimental probe of the absorptive coupling strength.
\item It vanishes when $\delta+\phi=n\pi$, independently of whether
    $\delta$ or $\phi$ individually vanish. In particular it survives
    for $\delta=0$, where all $CPT$ asymmetries vanish.
\item It is antisymmetric under flavour exchange: comparing
    $\nu_e\to\nu_\mu$ with $\nu_\mu\to\nu_e$ rather than neutrinos
    with antineutrinos.
\end{itemize}
Fig.~\ref{fig:DeltaT}, alongside $\Delta_{\rm CP}^{e\mu}$ for comparison.
  The two asymmetries share the same damping structure
  $e^{-\aabs t}|\sin(\Lambda t)|^2/|\Lambda|^2$ but differ in amplitude,
  making them complementary probes of the non-normal framework.

\section{Lepton Flavour Asymmetry}
\label{sec:application}

The $CP$ and $CPT$ asymmetries derived in Section~\ref{sec:asymmetries}
have a direct consequence for the lepton number content of the
active-neutrino sector. In this section we derive the individual
flavour asymmetries generated by the non-normal evolution and
establish an exact cancellation of the total lepton number that
is a structural consequence of the model.

\subsection{Purity structure and physical interpretation}

\begin{figure}[t]
  \centering
  \includegraphics[width=0.65\linewidth]{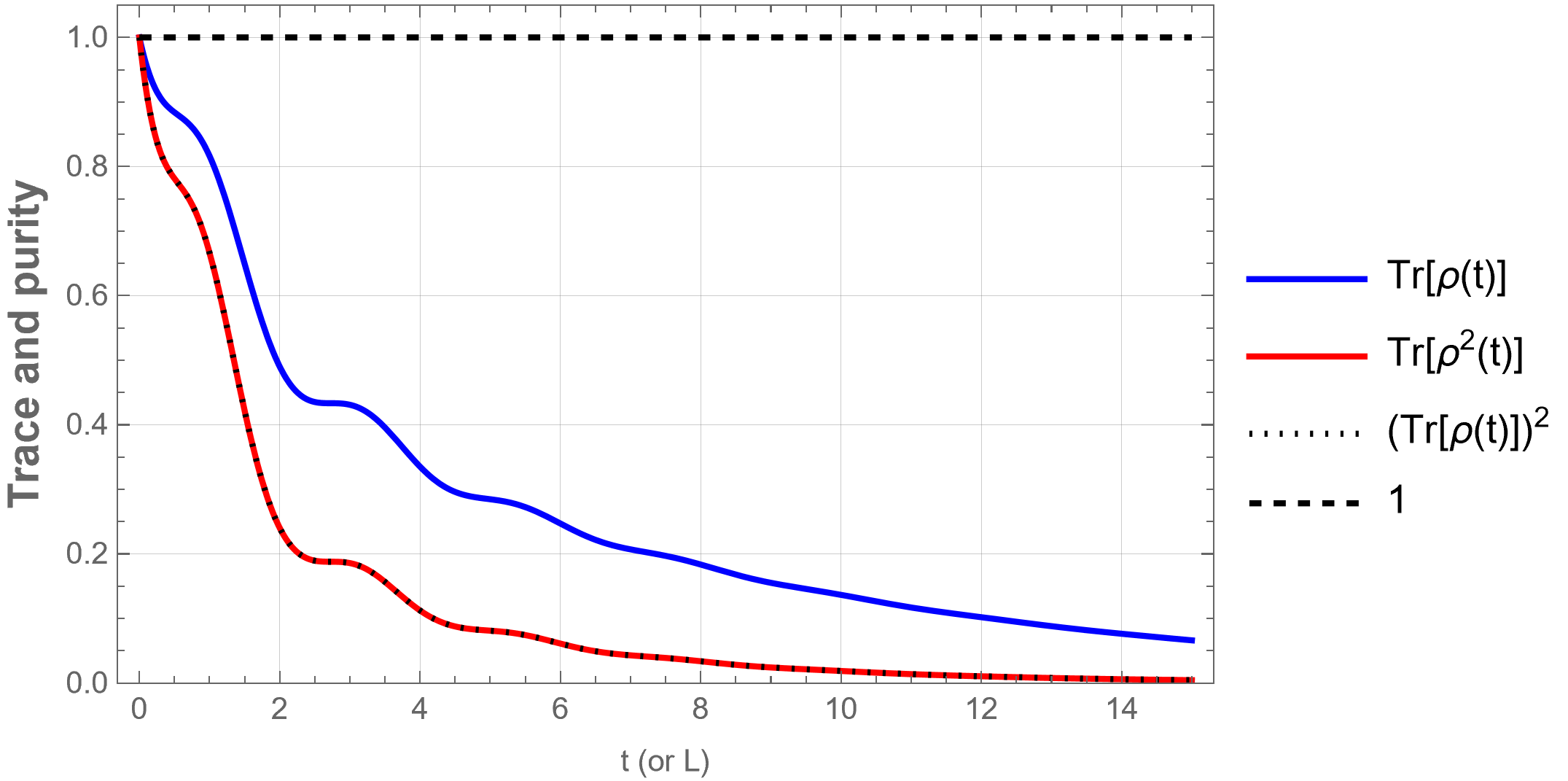}
\caption{Time evolution of $\mathrm{Tr}[\rho(t)]$ (blue),
$\mathrm{Tr}[\rho^2(t)]$ (red) and $(\mathrm{Tr}[\rho(t)])^2$
(dotted) for an initial $\nu_e$ state, for
$\theta=\phi=\delta=\pi/4$, $\Delta=-1$, $\lambda=1$,
$\aabs=0.5$. The red and dotted curves coincide at all times,
so the normalised purity
$P_{\rm norm}\equiv\mathrm{Tr}[\rho^2]/(\mathrm{Tr}[\rho])^2=1$
throughout, as expected for an initially pure state. Both
$\mathrm{Tr}[\rho]$ and $\mathrm{Tr}[\rho^2]$ decrease below the
pure-state reference (dashed), reflecting norm loss into unobserved
channels without quantum decoherence.}
  \label{fig:purity}
\end{figure}

The time evolution of the purity $\mathcal{P}(t)=\mathrm{Tr}[\rho^2(t)]$
provides a clean discrimination between physical mechanisms. Three
distinct cases arise:
\begin{itemize}
\item \textbf{Norm loss without decoherence}
    ($\mathrm{Tr}[\rho^2]=(\mathrm{Tr}[\rho])^2<1$):
    the state remains a pure-state projector with decreasing norm,
    corresponding to invisible decay or neutrino loss. The $L/E$
    scaling of the damping factor is the experimental signature.
\item \textbf{Decoherence without norm loss}
    ($\mathrm{Tr}[\rho]=1$, $\mathrm{Tr}[\rho^2]<1$):
    Lindblad evolution drives a genuine transition to a mixed state.
    The damping follows $e^{-\gamma L^n E^m}$ with exponents
    $n,m$ set by the microscopic mechanism.
\item \textbf{Combined} ($\mathrm{Tr}[\rho]<1$ and
    $\mathrm{Tr}[\rho^2]<(\mathrm{Tr}[\rho])^2$):
    both norm loss and coherence degradation are present simultaneously.
\end{itemize}
Since the non-Hermitian evolution generated by $H_{\rm eff}$
preserves the normalised purity
$P_{\rm norm}\equiv\mathrm{Tr}[\rho^2]/(\mathrm{Tr}[\rho])^2=1$,
our framework belongs to the first category: the system undergoes
norm loss without genuine quantum decoherence, as illustrated in
Fig.~\ref{fig:purity}. The lepton asymmetry derived below therefore originates from flavour-dependent norm loss that differs between neutrinos and antineutrinos, while the total norm loss is the same in both sectors.
This is a consequence of keeping the intrinsic background phase
$\delta$ invariant under $\nu\leftrightarrow\bar\nu$, while the mixing
phase $\phi$ changes sign. The non-commuting Hermitian and absorptive
parts of $H_{\rm eff}$ then translate this phase assignment into
different norm-loss rates, rather than into genuine coherence
degradation. 

\subsection{Total lepton number is exactly conserved}

First, we define the neutrino and antineutrino number densities as
\begin{align}
N_{\nu_\alpha} &= n_{\nu_\alpha}P_{\nu_\alpha\to\nu_\alpha}
    + n_{\nu_\beta}P_{\nu_\beta\to\nu_\alpha},\\
N_{\bar\nu_\alpha} &= n_{\bar\nu_\alpha}P_{\bar\nu_\alpha\to\bar\nu_\alpha}
    +n_{\bar\nu_\beta}P_{\bar\nu_\beta\to\bar\nu_\alpha},
\end{align}
where $n$ represents the initial number density per species.

Now, assuming an equal initial number density $n_0$ for all species,  as
appropriate for a thermal bath,
the total active-neutrino lepton number excess is given by
\begin{equation}
\Delta L(t)
= \sum_\alpha\bigl(N_{\nu_\alpha}(t) - N_{\bar\nu_\alpha}(t)\bigr)
= n_0\sum_\alpha
\bigl(\Delta_{\rm CPT}^{\alpha\alpha} + \Delta_{\rm CP}^{\beta\alpha}\bigr),
\qquad \beta\neq\alpha .
\end{equation}

We  will show that, under this assumption,  $\Delta L = 0$ exactly, with no further approximation. The total probability summed over all possible initial
and final flavours is
\begin{equation}
\sum_{\alpha,\beta} P_{\nu_\alpha\to\nu_\beta}(t)
= \sum_{\alpha,\beta} |A_{\beta\alpha}|^2
= \mathrm{Tr}(A^\dagger A),
\end{equation}
where in the last equality we have used
$\|A\|_F^2=\mathrm{Tr}(A^\dagger A)$.

Given that $A=R\,\mathcal{U}R^\dagger$, 
$RR^\dagger=R^\dagger R=\mathbb{I}$, and using the cyclic property
of the trace, the last expression simplifies to
\begin{equation}
\sum_{\alpha,\beta} P_{\nu_\alpha\to\nu_\beta}(t)
= \mathrm{Tr}(\mathcal{U}^\dagger \mathcal{U}).
\end{equation}

For antineutrinos, i.e.\ under the transformation $\phi\to-\phi$, we have
$R\to R^*$ and $\mathcal{U} \to \mathcal{U}$. Thus, taking into account that
$\widetilde{A}=R^* \mathcal{U} R^T$ and
$R^TR^*=R^*R^T=\mathbb{I}$, we obtain
\begin{equation}
\sum_{\alpha,\beta} P_{\bar\nu_\alpha\to\bar\nu_\beta}(t)
= \mathrm{Tr}(\tilde{A}^\dagger \tilde{A})
= \mathrm{Tr}(\mathcal{U}^\dagger \mathcal{U}).
\end{equation}

Therefore,
\begin{equation}
\sum_{\alpha,\beta} P_{\nu_\alpha\to\nu_\beta}
=
\sum_{\alpha,\beta} P_{\bar\nu_\alpha\to\bar\nu_\beta}
\quad\text{identically},
\label{eq:DLzero}
\end{equation}
giving $\Delta L = 0$ exactly. This result holds regardless of the
prescription for $\delta$, for any number of flavours, and for any
parameter values within our framework.

\subsection{Individual flavour asymmetries are non-zero}

Although $\Delta L = 0$, the individual flavour asymmetries need
not vanish. Define:
\begin{equation}
\Delta L_\alpha(t)
\equiv N_{\nu_\alpha}(t) - N_{\bar\nu_\alpha}(t)
= n_0\bigl(\Delta_{\rm CPT}^{\alpha\alpha}
  + \Delta_{\rm CP}^{\beta\alpha}\bigr),
\quad \alpha\neq\beta.
\end{equation}
For $\alpha=e$, inserting the explicit expressions from
Section~\ref{sec:asymmetries}, the contributions proportional
to $C_+|\sin(\Lambda t)|^2$ cancel exactly, leaving:
\begin{equation}
\Delta L_e(t)
= \frac{2n_0\lambda\sin 2\theta\,\sin\phi}
       {|\Lambda|^2}\,e^{-\aabs t}
\Bigl[A(t)\sin\delta
- \aabs\cos\delta\,|\sin(\Lambda t)|^2\Bigr],
\label{eq:DeltaLe}
\end{equation}
where $A(t)=k\sin(2\omega t)-\omega\sinh(2kt)$. By the same
calculation:
\begin{equation}
\Delta L_\mu(t) = -\Delta L_e(t),
\end{equation}
consistent with $\Delta L = 0$. The result
\begin{equation}
\Delta L_e = -\Delta L_\mu \neq 0
\end{equation}
is a genuine, non-trivial prediction of the framework: the
non-normal evolution redistributes lepton number between flavours
without creating or destroying it in total.
\begin{figure}[t]
  \centering
  \includegraphics[width=0.6\linewidth]{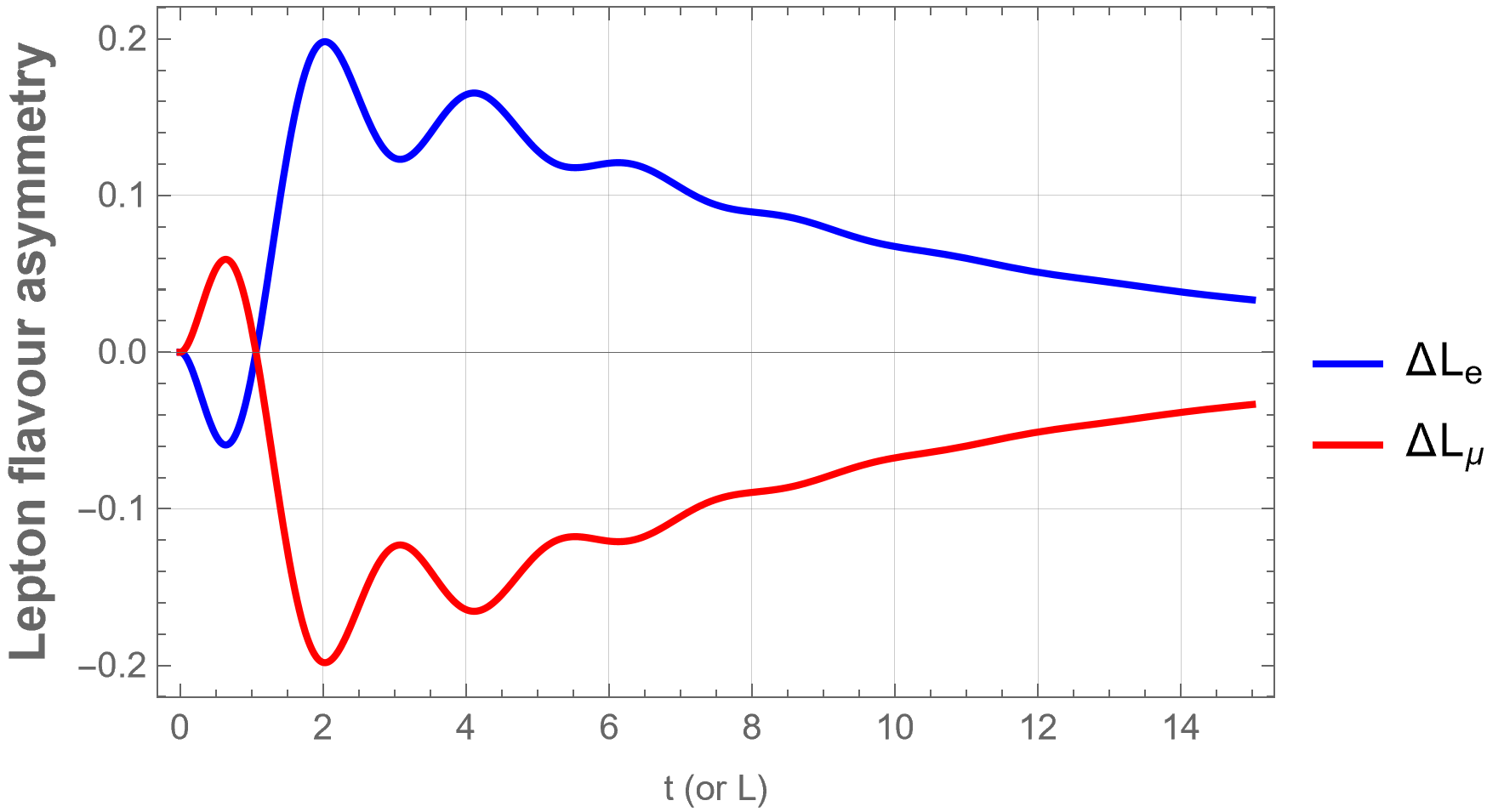}
\caption{Time evolution of the lepton flavour asymmetries
$\Delta L_e$ (blue) and $\Delta L_\mu=-\Delta L_e$ (red), for
$\theta=\phi=\delta=\pi/4$, $\Delta=-1$, $\lambda=1$,
$\aabs=0.5$. The flavour asymmetry is non-zero at maximal mixing,
whereas the total asymmetry $\Delta L=\Delta L_e+\Delta L_\mu$
vanishes identically. The transient character of $\Delta L_e$
reflects the net damping $e^{-(\aabs-2|k|)t}$.}
  \label{fig:DeltaLe}
\end{figure}
Figure~\ref{fig:DeltaLe} illustrates the time evolution of
$\Delta L_e$ and $\Delta L_\mu$ at maximal mixing; several
features of Eq.~\eqref{eq:DeltaLe} are worth noting:
\begin{itemize}
\item Both terms require $\aabs\neq 0$: the first through
    $k = \mathrm{Im}(\Lambda) \propto \aabs$ via the identity
    $2wk=\Delta\aabs$, and the second explicitly. The absorptive
    channel is therefore necessary for generating the flavour
    asymmetry, not merely for damping it.
\item $\Delta L_e$ does not vanish at maximal mixing $\theta=\pi/4$,
    where $\sin 2\theta = 1$. Moreover, the contributions
    proportional to $C_+$ cancel exactly in $\Delta L_e$ for any
    value of $\theta$, so the flavour asymmetry does not depend
    on $C_+$.
\item For non-zero $\lambda$, $\aabs$ and $\sin 2\theta$, the
    asymmetry is non-zero for generic $\delta$. It vanishes
    identically when $\phi = n\pi$, and otherwise only at the
    isolated times where
    $A(t)\sin\delta = \aabs\cos\delta\,|\sin(\Lambda t)|^2$.
\item Since $|\sin(\Lambda t)|^2$ and $\sinh(2kt)$ grow as
    $e^{2|k|t}$, the net envelope of $\Delta L_e$ is
    $e^{-(\aabs-2|k|)t}$, with $2|k|<\aabs$ for
    $\lambda\neq 0$. The asymmetry therefore has a transient
    character: it builds up over a timescale
    $\sim\mathrm{Re}(\Lambda)^{-1}$ and is attenuated over
    $\sim(\aabs-2|k|)^{-1}$, which is longer than
    $\aabs^{-1}$.
\end{itemize}

\subsection{Physical interpretation}

The non-zero flavour asymmetry $\Delta L_e=-\Delta L_\mu$ has a
transparent physical origin. The absorptive term acts on $\nu_2$
in the mass basis. Because $[H_{\rm herm},\Gamma]\neq 0$, the
decaying modes of $H_{\rm eff}$ are non-orthogonal combinations of
mass eigenstates. Their projection onto flavour through $R$ depends
on $\phi$, which is reversed for antineutrinos while $\delta$ is
not. As a result, the electron and muon components lose norm at
different rates in the two sectors. Since
$\mathrm{Im}\,E_+ + \mathrm{Im}\,E_- = -\aabs$ for all parameter
values, the total absorption rate is fixed; the parameters only
determine how it is shared between the two effective modes, from
$(0,\aabs)$ in the selective-damping limit to
$(\aabs/2,\aabs/2)$ in the quasi-global regime. The result is
a redistribution of lepton number between flavours, even though the
total is conserved.

This is a qualitatively different outcome from standard leptogenesis
scenarios, which require $\Delta L \neq 0$ as the source of $\Delta B$.
Here the non-normal dynamics produces instead a pure flavour asymmetry.
Whether this has cosmological consequences depends on physics beyond
the present two-flavour framework. In the full three-flavour theory,
interactions that are in equilibrium at different temperatures ---
such as the tau and muon Yukawa couplings --- treat $L_e$, $L_\mu$,
and $L_\tau$ differently, and a state with $\Delta L_e \neq 0$ can
in principle source observable effects even when $\Delta L = 0$.
A systematic investigation of these effects in the three-flavour
extension, including the possible role of the T-asymmetry
$\Delta_T^{e\mu}\neq 0$ derived in Section~\ref{sec:asymmetries},
is left for future work.

\section{Conclusions and outlook}
\label{sec:conclusions}
We have studied two-flavour neutrino oscillations in which a coherent
off-diagonal interaction, with strength $\lambda$ and phase $\delta$,
coexists with an absorptive channel acting on a single mass eigenstate.
Whenever both are present, $[H_{\rm herm},\Gamma]\neq 0$ and the effective
Hamiltonian is non-normal, so coherent mixing and absorption cannot be
factorised as in the normal case.

On the formal side, we presented two independent approaches to the
non-Hermitian dynamics: a non-trace-preserving Liouville equation
corresponding to the no-jump sector of a Lindblad evolution, and
biorthogonal diagonalisation of the non-Hermitian Hamiltonian. For our
two-flavour model, we then derived the exact evolution operator in closed
form using the Cayley--Hamilton theorem. Depending on the relative
size of $\Delta$, $\lambda$ and $\aabs$, the evolution reduces to quasi-global
damping, to standard invisible decay of one mass eigenstate, or, at the
critical point $2\lambda=\aabs$ with $\Delta=\delta=0$, to
polynomial-times-exponential evolution. For $\Delta\neq 0$ the evolution is
bounded for any real $\lambda$ and $\aabs\geq 0$. The asymmetries are
damped by the envelope $e^{-(\aabs-2|k|)t}$, with $k={\rm Im}\,\Lambda$,
rather than by $e^{-\aabs t}$ alone.

For initially pure states the no-jump evolution maps pure states into
unnormalised pure states, so $P_{\rm norm}={\rm Tr}[\rho^2]/({\rm Tr}\rho)^2=1$
throughout while ${\rm Tr}\,\rho$ decreases. The framework therefore
describes norm loss into unobserved channels, not quantum decoherence. The
two can be distinguished through purity-sensitive observables and through
the energy--baseline scaling of the damping.

We obtained closed-form $CP$, $CPT$ and $T$ asymmetries for all flavour
channels by comparing neutrinos and antineutrinos propagating in the same
fixed background. The mixing phase $\phi$ is conjugated, while the phase
$\delta$, fixed by the selected background configuration in analogy with
background coefficients in the Standard Model Extension, is not. This
comparison is not a $CPT$ transformation of the complete system: the
resulting $CPT$ asymmetries are induced by the background, in the same
sense as matter-induced asymmetries. They exist only because of the
differential phase assignment; if $\delta$ were also reversed they would
vanish identically, and the $CP$ asymmetries would reduce to the $T$
asymmetry. The $T$ asymmetry,
$\Delta_T^{e\mu}\propto \aabs\lambda\sin 2\theta\,\sin(\delta+\phi)$,
is linear in the absorptive coupling.  Since $P_{e\mu}=P_{\mu e}$ for any unitary two-flavour evolution, this
asymmetry is a direct signature of the non-normal structure, and,
unlike the $CPT$ asymmetries, it does not depend on the phase
prescription. For non-zero $\lambda$, $\aabs$
and $\sin 2\theta$ it vanishes only when $\delta+\phi=n\pi$, and in
particular it survives at $\delta=0$, where all $CPT$ asymmetries vanish.
Since $\Delta_T^{e\mu}=\Delta_{\rm CP}^{e\mu}-\Delta_{\rm CPT}^{\mu e}$, it
is not independent of the $CP$ and $CPT$ asymmetries, but it isolates the
absorptive coupling in a single observable.

For equal initial populations the net neutrino--antineutrino asymmetry
vanishes identically, $\Delta L=0$. This follows from the invariance of
${\rm Tr}(\mathcal{U}^\dagger \mathcal{U})$ under $R\to R^*$. It holds for any number of
flavours and for either prescription for $\delta$, and it does not rely on
non-normality. The flavour asymmetries do not vanish:
$\Delta L_e=-\Delta L_\mu$ has a closed form, is non-zero at maximal mixing,
and requires $\lambda\neq 0$, $\aabs\neq 0$ and $\sin\phi\neq 0$. It is
therefore a direct consequence of the non-normal structure. The total norm
decreases equally in the neutrino and antineutrino sectors, but its
distribution among flavours differs between them.

Whether such a flavour asymmetry has cosmological consequences is left
open. Electroweak sphalerons conserve $B/3-L_\alpha$, and at leading order
the final baryon asymmetry is fixed by the total $B-L$. A pure flavour
asymmetry with $\Delta L=0$ therefore does not by itself source a baryon
asymmetry. A net effect would require additional flavour-dependent
processes, such as washout acting differently on the individual lepton
flavours, or unequal initial populations. Assessing this requires the
three-flavour extension of the framework, together with a concrete
realisation of the background field in the early Universe, which we leave
for future work.

\section*{Acknowledgments}
GB is supported by  PID2023-151418NB-I00 funded by MCIU/AEI/10.13039/501100011033/, 
and by the European ITN project HIDDeN (H2020-MSCA-ITN-2019/860881-HIDDeN). A. M. G. acknowledges funding by the Dirección de Gestión de la Investigación at PUCP, through grant
DFI-PUCP-PI1144.

\appendix

\section{Useful Identities}
\label{app:identities}

\subsection{Complex trigonometric identities}

For $\Lambda=w+ik$:
\begin{align}
|\sin(\Lambda t)|^2 &= \tfrac{1}{2}[\cosh(2kt)-\cos(2wt)],\\
|\cos(\Lambda t)|^2 &= \tfrac{1}{2}[\cosh(2kt)+\cos(2wt)],\\
\sin(\Lambda^* t)\cos(\Lambda t) &= \tfrac{1}{2}[\sin(2wt)-i\sinh(2kt)],\\
\frac{\Delta+i\frac{\aabs}{2}}{w+ik}
    &= \frac{\Delta w+k\frac{\aabs}{2}}{|\Lambda|^2}
    + i\frac{\frac{\aabs}{2}w-\Delta k}{|\Lambda|^2},\qquad
\frac{1}{\Lambda^*} = \frac{w+ik}{|\Lambda|^2}.
\end{align}
\textbf{Important relation:} Equating imaginary parts of $\Lambda^2$
gives $2wk=\Delta\aabs$, valid for all parameter values.

\subsection{List of useful matrix element identities}

\begin{enumerate}
\item $|u_{12}|^2=|u_{21}|^2
    = e^{-\aabs t}\dfrac{\lambda^2}{|\Lambda|^2}|\sin(\Lambda t)|^2$.

\item $\mathrm{Re}(u_{12}^*u_{21})=|u_{21}|^2\cos 2\delta$ and
    $\mathrm{Im}(u_{12}^*u_{21})=-|u_{21}|^2\sin 2\delta$.

\item $\mathrm{Re}[u_{11}(u_{21}-u_{12})^*]
    = -e^{-\aabs t}\dfrac{\lambda\sin\delta}{|\Lambda|^2}
    \Bigl[w\sin(2wt)+k\sinh(2kt)
    +\tfrac{\aabs}{2}(2|\sin(\Lambda t)|^2)\Bigr]$.

\item $\mathrm{Im}[u_{11}(u_{21}-u_{12})^*]
    = -e^{-\aabs t}\dfrac{\lambda\sin\delta}{|\Lambda|^2}
    \Bigl[k\sin(2wt)-w\sinh(2kt)
    -\Delta(2|\sin(\Lambda t)|^2)\Bigr]$.

\item $\mathrm{Re}[u_{22}(u_{21}-u_{12})^*]
    = -e^{-\aabs t}\dfrac{\lambda\sin\delta}{|\Lambda|^2}
    \Bigl[w\sin(2wt)+k\sinh(2kt)
    -\tfrac{\aabs}{2}(2|\sin(\Lambda t)|^2)\Bigr]$.

\item $\mathrm{Im}[u_{22}(u_{21}-u_{12})^*]
    = -e^{-\aabs t}\dfrac{\lambda\sin\delta}{|\Lambda|^2}
    \Bigl[k\sin(2wt)-w\sinh(2kt)
    +\Delta(2|\sin(\Lambda t)|^2)\Bigr]$.

    \item \begin{align*}
    |u_{11}|^2+|u_{12}|^2 &=e^{-\aabs t}\Bigg[|\cos({\Lambda t})|^2 + \dfrac{(\Delta w +\dfrac{\aabs}{2}k)}{|\Lambda|^2}
     \sinh{(2kt)} \\
     &+\dfrac{(\dfrac{\aabs}{2} w-\Delta k)}{|\Lambda|^2}
     \sin{(2wt)} +
    \dfrac{(\Delta^2+\dfrac{\aabs^2}{4}+\lambda^2)}{|\Lambda|^2} |\sin({\Lambda t})|^2
    \Bigg]
    \end{align*}
    
    \item \begin{align*}
    |u_{22}|^2+|u_{12}|^2 &=e^{-\
\aabs t}\Bigg[|\cos({\Lambda t})|^2 -\dfrac{(\Delta w +\dfrac{\aabs}{2}k)}{|\Lambda|^2}
     \sinh{(2kt)} \\
     &-\dfrac{(\dfrac{\aabs}{2} w-\Delta k)}{|\Lambda|^2}
     \sin{(2wt)} +
    \dfrac{(\Delta^2+\dfrac{\aabs^2}{4}+\lambda^2)}{|\Lambda|^2} |\sin({\Lambda t})|^2
    \Bigg]
    \end{align*}

\item $\mathrm{Re}(u_{11}u_{22}^*)
    = e^{-\aabs t}\!\left[|\cos(\Lambda t)|^2
    -\frac{\Delta^2+\frac{\aabs^2}{4}}{|\Lambda|^2}
    |\sin(\Lambda t)|^2\right]$.

\item $\mathrm{Re}[u_{11}(u_{12}+u_{21})^*]
    = e^{-\aabs t}\dfrac{\lambda\cos\delta}{|\Lambda|^2}
    \Bigl[-k\sin(2wt)+w\sinh(2kt)
    +\Delta(2|\sin(\Lambda t)|^2)\Bigr]$.

\item $\mathrm{Im}[u_{11}(u_{12}+u_{21})^*]
    = e^{-\aabs t}\dfrac{\lambda\cos\delta}{|\Lambda|^2}
    \Bigl[w\sin(2wt)+k\sinh(2kt)
    +\tfrac{\aabs}{2}(2|\sin(\Lambda t)|^2)\Bigr]$.

\item $\mathrm{Re}[u_{22}(u_{12}+u_{21})^*]
    = e^{-\aabs t}\dfrac{\lambda\cos\delta}{|\Lambda|^2}
    \Bigl[-k\sin(2wt)+w\sinh(2kt)
    -\Delta(2|\sin(\Lambda t)|^2)\Bigr]$.

\item $\mathrm{Im}[u_{22}(u_{12}+u_{21})^*]
    = e^{-\aabs t}\dfrac{\lambda\cos\delta}{|\Lambda|^2}
    \Bigl[w\sin(2wt)+k\sinh(2kt)
    -\tfrac{\aabs}{2}(2|\sin(\Lambda t)|^2)\Bigr]$.

\item
\begin{align*}
\mathrm{Re}[u_{22}u_{12}^*]
= \frac{e^{-\aabs t}\lambda}{2|\Lambda|^2}&
\Bigl[\bigl(-k\sin(2wt)+w\sinh(2kt)
    -\Delta(2|\sin(\Lambda t)|^2)\bigr)\cos\delta\\
&+\bigl(w\sin(2wt)+k\sinh(2kt)
    -\tfrac{\aabs}{2}(2|\sin(\Lambda t)|^2)\bigr)\sin\delta\Bigr].
\end{align*}

\item
\begin{align*}
\mathrm{Im}[u_{22}u_{12}^*]
= -\frac{e^{-\aabs t}\lambda}{2|\Lambda|^2}&
\Bigl[\bigl(-k\sin(2wt)+w\sinh(2kt)
    -\Delta(2|\sin(\Lambda t)|^2)\bigr)\sin\delta\\
&-\bigl(w\sin(2wt)+k\sinh(2kt)
    -\tfrac{\aabs}{2}(2|\sin(\Lambda t)|^2)\bigr)\cos\delta\Bigr].
\end{align*}
\end{enumerate}

\section{Projection of the non-Hermitian Liouvillian onto the Pauli basis}
\label{app:PauliLiouvillian}

\appendix

In this appendix we derive Eq.~\eqref{LPmatrixelements}, which gives the
matrix representation of the non-Hermitian Liouvillian in the Pauli
operator basis. Starting from the Pauli expansion
\begin{equation}
\rho(t)
=
\frac{1}{2}
\sum_{\nu=0}^{3}
\varrho_{\rm P}^{\nu}(t)\sigma_\nu ,
\end{equation}
and using the linearity of \(\mathcal L_{\rm NH}\), one obtains
\begin{equation}
\mathcal L_{\rm NH}[\rho(t)]
=
\frac{1}{2}
\sum_{\nu=0}^{3}
\varrho_{\rm P}^{\nu}(t)
\mathcal L_{\rm NH}[\sigma_\nu].
\end{equation}
On the other hand,
\begin{equation}
\dot{\rho}(t)
=
\frac{1}{2}
\sum_{\mu=0}^{3}
\dot{\varrho}_{\rm P}^{\mu}(t)\sigma_\mu .
\end{equation}
Now, considering that:
\begin{equation}
\dot{\rho}(t)
=
\mathcal L_{\rm NH}[\rho(t)],
\end{equation}
we have
\begin{equation}
\frac{1}{2}
\sum_{\mu=0}^{3}
\dot{\varrho}_{\rm P}^{\mu}(t)\sigma_\mu
=
\frac{1}{2}
\sum_{\nu=0}^{3}
\varrho_{\rm P}^{\nu}(t)
\mathcal L_{\rm NH}[\sigma_\nu].
\end{equation}
Multiplying by \(\sigma_\alpha\), taking the trace, and using
\begin{equation}
{\rm Tr}(\sigma_\alpha\sigma_\mu)
=
2\delta_{\alpha\mu},
\end{equation}
one finds
\begin{equation}
\dot{\varrho}_{\rm P}^{\alpha}(t)
=
\frac{1}{2}
\sum_{\nu=0}^{3}
\varrho_{\rm P}^{\nu}(t)
{\rm Tr}\!\left[
\sigma_\alpha
\mathcal L_{\rm NH}[\sigma_\nu]
\right].
\end{equation}
Comparing the above expression with
\begin{equation}
\dot{\varrho}_{\rm P}^{\alpha}
=
\sum_{\nu=0}^{3}
(\mathbb L_{\rm P})_{\alpha\nu}
\varrho_{\rm P}^{\nu},
\end{equation}
we obtain
\begin{equation}
(\mathbb L_{\rm P})_{\alpha\nu}
=
\frac{1}{2}
{\rm Tr}\!\left[
\sigma_\alpha
\mathcal L_{\rm NH}[\sigma_\nu]
\right].
\end{equation}


\bibliographystyle{unsrt}
\bibliography{references}
\end{document}